\documentclass[usenatbib]{mn2e}   
\pdfoutput=1 
\usepackage{graphicx} %\documentstyle[psfig,amssymb]{mn} 
\usepackage{amsmath}
\usepackage{url}

\title[The AGN XLF at $z=3-5$]{The X-ray luminosity function of Active Galactic
  Nuclei in the redshift interval {\it z}\,=\,3\,--\,5.}
\author[Georgakakis et al.]   {A. Georgakakis$^{1, 2}$, J. Aird$^3$, J. Buchner$^1$, 
    M. Salvato$^1$, M.-L. Menzel$^1$,  W. N. Brandt$^4$, 
    \newauthor I. D. McGreer$^{5}$, T. Dwelly$^{1}$, G. Mountrichas$^2$, 
    K. Koki$^{2,6}$, I. Georgantopoulos$^{2}$, L.-T. Hsu$^{1}$,
    \newauthor A. Merloni$^1$, 
    Z. Liu$^{1,7}$,
    K. Nandra$^1$, N. P. Ross$^{8,9}$\\ \\ 
\\ \\ 
 $^1$Max Planck  Institut f\"{u}r Extraterrestrische Physik, Giessenbachstra\ss e 1, D-85748, Garching bei M\"{u}nchen, Germany\\
 $^2$IAASARS, National Observatory of Athens, GR-15236 Penteli, Greece\\
 $^3$Institute of Astronomy, University of Cambridge, Madingley Road, Cambridge, CB3 0HA\\
 $^4$Department of Astronomy and Astrophysics and the Institute for Gravitation and the Cosmos,
 The Pennsylvania State University,\\ 525 Davey Lab, University Park, PA 16802, USA\\
 $^5$Steward Observatory, The University of Arizona, 933 North Cherry Avenue, Tucson, AZ 85721-0065, USA\\ 
 $^6$Department of Statistics, Athens University of Economics and
 Business, Athens, 13476, Greece\\
 $^7$National Astronomical Observatories,  Chinese Academy of Sciences, Beijing 100012, People's Republic of China\\
 $^8$Department of Physics, Drexel University, 3141 Chestnut Street, Philadelphia, PA
19104, USA\\
 $^9$Institute for Astronomy, University of Edinburgh, Royal Observatory, Edinburgh EH9
3HJ, UK.
}

\begin{document}
\maketitle

\begin{abstract} We  combine deep X-ray  survey data from  the Chandra
observatory and  the wide-area/shallow  XMM-XXL field to  estimate the
AGN  X-ray luminosity  function in  the redshift  range  $z=3-5$.  The
sample consists of nearly 340 sources with either photometric (212) or
spectroscopic (128)  redshift in the above range.   The combination of
deep and shallow survey fields  also provides a luminosity baseline of
three orders of  magnitude, $L_X( \rm 2 - 10  \, keV)\approx 10^{43} -
10^{46} \, erg \, s^{-1}$ at  $z>3$.  We follow a Bayesian approach to
determine the binned AGN space  density and explore their evolution in
a  model-independent  way.   Our  methodology  properly  accounts  for
Poisson  errors  in  the   determination  of  X-ray  fluxes  and
uncertainties in photometric  redshift estimates.  We demonstrate that
the latter  is essential for unbiased measurement  of space densities.
We  find  that the  AGN  X-ray  luminosity  function evolves  strongly
between  the redshift intervals  $z=3-4$ and  $z=4-5$.  There  is also
suggestive evidence that the amplitude of this evolution is luminosity
dependent. The  space density of  AGN with $L_X(  \rm 2 - 10  \, keV)<
10^{45} \, erg \, s^{-1}$ drops  by a factor of 5 between the redshift
intervals above,  while the  evolution of brighter  AGN appears  to be
milder.   Comparison of  our X-ray  luminosity function  with  that of
UV/optical selected QSOs at similar redshifts shows broad agreement at
bright luminosities,  $L_X( \rm  2 - 10  \, keV)  > 10^{45} \,  erg \,
s^{-1}$.   At fainter  luminosities X-ray  surveys measure  higher AGN
space  densities.   The  faint-end  slope   of  UV/optical  luminosity
functions  however, is  steeper than  for X-ray  selected  AGN.  This
implies  that  the  type-I  AGN  fraction  increases  with  decreasing
luminosity at $z>3$, opposite to trends established at lower redshift.
We also assess the significance of AGN in keeping the hydrogen ionised
at  high  redshift.  Our  X-ray  luminosity  function yields  ionising
photon  rate densities  that  are insufficient  to  keep the  Universe
ionised  at redshift  $z>4$. A  source  of uncertainty  in this
calculation is  the escape fraction  of UV photons for  X-ray selected
AGN.
\end{abstract}

\section{Introduction}\label{sec_intro}  In recent  years observations
have  established that  supermassive  black holes  (SMBHs) are  nearly
ubiquitous      in      local     spheroids      \citep{Magorrian1998,
Kormendy_Ho2013}. These  relic black holes are believed  to have grown
their masses  at earlier times  mostly via accretion of  material from
larger scales \citep[e.g.][]{Soltan1982, Marconi2004}.  Questions that
remain open are when during  the lifetime of the Universe these events
occurred  and under what  physical conditions  black holes  grow their
masses.   Moreover,  observations  show  that in  the  local  Universe
correlations exist between the mass of SMBHs and the properties of the
stellar component of the bulges in which they reside, such as velocity
dispersion  \citep[e.g][]{Ferrarese2000,  Gebhardt2000,  Gultekin2009,
Graham2011},        luminosity        \citep[e.g][]{McLure_Dunlop2002,
Marconi_Hunt2003,         Gultekin2009},         dynamical        mass
\citep[e.g.][]{Magorrian1998,     Marconi_Hunt2003,    Haring_Rix2004,
Graham2012}  and  central  light concentration  \citep[][]{Graham2001,
Savorgnan2013}.  Such  correlations suggest a link  between the growth
of black holes and the  formation of their host galaxies, although the
exact  nature of  such  an  interplay is  still  not well  understood.
Processes that  can establish such  correlations include a  common gas
reservoir that  both feeds the central  black hole and  forms stars on
larger  scales, outflows  related to  the energy  output from  the AGN
itself that affect the Inter-Stellar Medium and regulate the formation
of    stars   \citep{Silk1998,    Fabian1999,    King2003,   King2005,
DiMatteo2005,  Croton2006}, or  the  merging history  of galaxies  and
their SMBHs \citep[e.g.][]{Jahnke2011}.

One approach for improving our understanding of the formation of SMBHs
as a function of cosmic time and their relation to their host galaxies
is population studies of  Active Galactic Nuclei (AGN), which signpost
accretion  events onto  SMBHs.   This  requires a  census  of the  AGN
population  across redshift  to constrain  for example,  the accretion
history of the  Universe or study the incidence  of active black holes
among galaxies.   In that respect,  the AGN luminosity  function, i.e.
their comoving space  density as a function of  redshift and accretion
luminosity, is one of the fundamental quantities that characterise the
demographics  of active  black  holes.  The  cosmic  evolution of  AGN
leaves  imprints  on  the  shape  and  overall  normalisation  of  the
luminosity function.   The total mass density locked  into black holes
at different epochs  can be inferred by direct  integration of the AGN
luminosity function, under  assumptions about the radiative efficiency
of  the accretion  process and  after applying  appropriate bolometric
luminosity corrections \citep[e.g][]{Marconi2004, Aird2010, Ueda2014}.
The  space density of  AGN split  by host  galaxy properties,  such as
stellar mass, morphology or level of star-formation, provides clues on
the  interplay  between  black  hole accretion  and  galaxy  evolution
\citep{Georgakakis2009,   Georgakakis2011,   Aird2012,  Bongiorno2012,
Georgakakis2014}.

Selection   at  UV/optical   wavelengths  \citep[e.g.][]{Richards2009,
Ross2012} currently provides the largest spectroscopic AGN samples for
luminosity   function   calculations  \citep[e.g.][]{Ross2013}.    The
downside is that the UV/optical  continuum of AGN is sensitive to dust
extinction along  the line--of--sight and dilution by  the host galaxy
at  faint accretion luminosities.   Observations at  X-ray wavelengths
can mitigate  these issues \citep[e.g][]{Brandt_Alexander2015}.  X-ray
photons,  particularly  at  rest-frame  energies  $\rm  >2\,keV$,  can
penetrate nearly unaffected large  columns of intervening gas and dust
clouds  ($\rm N_H  \ga 10^{22}\,cm^{-2}$),  thereby  providing samples
least affected  by obscuration  biases.  Moreover, the  X-ray emission
associated with  stellar processes is typically 2  orders of magnitude
fainter than  the AGN radiative output and  therefore contamination or
dilution effects  by the host galaxy  are negligible at  X-rays over a
wide baseline  of accretion luminosities.  X-ray  surveys also benefit
from  a well  defined selection  function that  is relatively  easy to
quantify and account  for in the analysis.  The  disadvantage of X-ray
selection  is that  the detected  AGN  are often  optically faint  and
therefore    spectroscopic    follow-up    observations   are    often
expensive. Nevertheless, intensive multiwavelength campaigns in recent
years substantially  increased the number of X-ray  survey fields with
sufficient   quality  ancillary   data  for   reliable   X-ray  source
identification and redshift  measurements using either spectroscopy or
photometric  methods.   Early   results  by  \cite{Cowie2003}  on  the
redshift evolution of the  X-ray AGN space density and \cite{Ueda2003}
on the hard band (2-10\,keV) X-ray luminosity function and obscuration
distribution of AGN have been  expanded recently both in terms of data
and  analysis  methodology  \citep{Yencho2009,  Ebrero2009,  Aird2010,
Burlon2011,  Ueda2014, Buchner2015,  Aird2015,  Miyaji2015}.  Although
important details  on the shape  of the X-ray luminosity  function and
the obscuration distribution of AGN are still debated \citep{Ueda2014,
Buchner2015, Aird2015, Miyaji2015}, the overall evolution of the X-ray
luminosity  function  is  reasonably  well  constrained  at  least  to
$z\approx3$.  The initial increase of the AGN X-ray luminosity density
from  $z=0$ to  $z\approx1.5$ is  followed by  a broad  plateau  up to
$z\approx2.5-3$  and  a  decline  at higher  redshift.   However,  the
amplitude  of the X-ray  AGN evolution  at $z\ga3$  is still  not well
constrained.  Early  studies suggested a  moderate decline of  the AGN
space  density  at  $z>3$  \citep{Yencho2009,  Ebrero2009,  Aird2010},
contrary  to claims  for  a rapid  drop \citep{Brusa2009,  Civano2011,
Vito2013, Kalfountzou2014} that can  be parametrised by an exponential
law  in redshift \citep{Gilli2007}  similar to  the optical  QSO space
density evolution \citep{Schmidt1995,  Richards2006}.  Central to this
debate is  the typically  small X-ray AGN  sample sizes at  $z>3$. For
example,  there  are 209  and  141  X-ray  AGN with  spectroscopic  or
photometric redshifts  above $z=3$ in the most  recent compilations of
\cite{Kalfountzou2014}   and   \cite{Vito2014}  respectively.    These
numbers should be  compared with sample sizes of  few thousands AGN at
$z<3$  for   the  most   recent  X-ray  luminosity   function  studies
\citep[e.g][]{Ueda2014, Buchner2015, Aird2015, Miyaji2015}

 Better constraints on the  form and amplitude of the evolution of
X-ray AGN at $z>3$ will have implications for the contribution of this
population to the  UV photon field density that is  needed to keep the
Universe  ionised at high  redshift. \cite{Haardt_Madau2012}  used the
\cite{Ueda2003}  X-ray  luminosity  function  to  predict  a  moderate
contribution of AGN to the hydrogen ionising radiation field at $z>3$,
in broad  agreement with constraints derived  from UV/optical selected
QSO luminosity  functions \citep[e.g.][]{Fontanot2007, Masters2012, McGreer2013} and
other  X-ray AGN  studies \citep{Barger2003_hiz,  Grissom2014}.  There
are also claims however, that AGN provide an important contribution to
the   photoionisation   rate   at  high   redshift   \citep{Fiore2012,
Glikman2011, Giallongo2015}. This  discrepancy emphasises the need for
further work to improve measurements  of the AGN space density at high
redshift and to  better understand their role in  the re-ionisation of
the Universe.

In this paper we combine deep Chandra and wide-area/shallow XMM-Newton
survey fields to compile one  of the largest samples of X-ray selected
AGN  at $z=3-5$  to  date.   A Bayesian  methodology  is developed  to
correctly  account  for  photometric  redshift  uncertainties  and  to
determine in  a non-parametric way  the AGN comoving space  density in
the redshift intervals $z=3-4$ and $z=4-5$ and over 3 decades in X-ray
luminosity   [$\log  L_X(\rm   2-10\,keV)  \approx   43-46$   in  $\rm
erg\,s^{-1}$].  Although  parametric models are also fit  to the data,
we emphasise  the importance of non-parametric  estimates to determine
in  a model-independent  way the  shape and  overall evolution  of the
X-ray luminosity function.  Throughout  this paper we adopt $\rm H_{0}
= 70  \, km \,  s^{-1} \, Mpc^{-1}$,  $\rm \Omega_{M} = 0.3$  and $\rm
\Omega_{\Lambda} = 0.7$.

\section{Data} 

For the determination of the X-ray luminosity function in the redshift
interval $z=3-5$ we combine  Chandra and XMM-Newton X-ray surveys with
different characteristics  in terms of area coverage  and X-ray depth.
These are the 4\,Ms Chandra Deep Field South \citep[CDFS;][]{Xue2011, 
Rangel2013},     the     2\,Ms     Chandra    Deep     Field     North
\citep[CDFN;][]{Alexander2003, Rangel2013},  the Extended Groth Strip
International  Survey  field  \citep[AEGIS,][]{Davis2007, Laird2009,
Nandra2015},    the     Extended    Chandra    Deep     Field    South
\citep[ECDFS;][]{Lehmer2005},         the    Chandra     COSMOS         field
\citep[C-COSMOS,][]{Elvis2009} and the equatorial field of the XMM-XXL 
survey.

\begin{table*}
\caption{Number of X-ray sources in the full-band selected sample}\label{table_data}

\begin{tabular}{l c cccc  cccc}

%\hline
%\multicolumn{9}{c}{Full-band selected sample.} \\
\hline
field & solid angle   & Number of      & Number of            & \multicolumn{3}{c}{photometric redshift}       &  \multicolumn{3}{c}{spectroscopic redshift} \\
      &  ($\rm arcmin^2$)    & X-ray sources  & optical/infrared IDs &   full sample  &  $3<z<4$ & $4<z<5$ &   full sample  &  $3<z<4$ & $4<z<5$ \\

 (1) & (2) & (3) & (4) & (5) & (6) & (7) & (8) & (9) & (10) \\
\hline

4Ms CDFS  & 271.4    & 422   & 418  & 151  & 16  & 3  &  268  & 9 & 1 \\ 

CDFN      & 412.9    & 453   & 436   &  171    &  11   & 5   & 265      &  9  & 2 \\

ECDFS     & 643.0    &  407   &  399  &  267    & 30    & 11   &  134     &  3  & 1 \\

AEGIS-XD  & 934.6    &  818   & 804  & 478  & 37  & 13 &  326  & 16 & 0 \\ 

AEGIS-W  &  724.3    &  415   & 413  & 267  & 26  & 8 &  146  & 2 & 1 \\ 

COSMOS    & 3258.0    & 1435  & 1393 & 635  & 38  & 14 &  758  & 21 & 4 \\ 

XMM-XXL   & 64728.0    & 7493  & 3798 &  --  &  -- & -- & 2553  & 54 & 5\\ 

Total     & 70972.2    &  --  &  --  & -- & 158  & 54 &  --   & 114& 14 \\
\hline

\end{tabular}
\begin{list}{}{}
\item  (1) Name  of  the X-ray survey  fields  used in  this
work.  The last  row  lists the  total  number of  sources for  the
combined fields. For that last row only the columns that correspond to
the total number of sources with photometric or spectroscopic redshift
in  the intervals  $3<z<4$ and  $4<z<5$ are  listed.  This  is because
these  are the  relevant numbers  to  the analysis  presented in  this
paper.   (2) Solid  angle of  each sample  in square  arcminutes after
excluding  regions of poor/no  photometry. (3)  Total number  of X-ray
sources  detected in the  0.5-7\,keV energy  band in  the case  of the
Chandra, or the 0.5-8\,keV band for the XMM-XXL sample.  (4) Number of
full-band detected  X-ray sources with  optical/infrared associations.
(5)  Number of  sources with  photometric redshift  estimates  in each
survey  field.    Photometric  redshifts  for  the   XMM-XXL  are  not
available.  (6)  Number of sources  with photometric redshifts  in the
interval  $z=3-4$.    This  the   sum  of  the   photometric  redshift
Probability Distribution Functions rounded to the nearest integer.  No
photometric redshift  estimates are available for  the XMM-XXL sample.
(7) Same  as column  (6) but for  the redshift interval  $z=4-5$.  (8)
Number of sources with spectroscopic redshift estimates in each survey
field.   (9) Number  of sources  with spectroscopic  redshifts  in the
interval $z=3-4$.  (10) Number of sources with spectroscopic redshifts
in the interval $z=4-5$.
\end{list}
\end{table*}

\begin{figure*}
\begin{center}
\includegraphics[height=0.85\columnwidth]{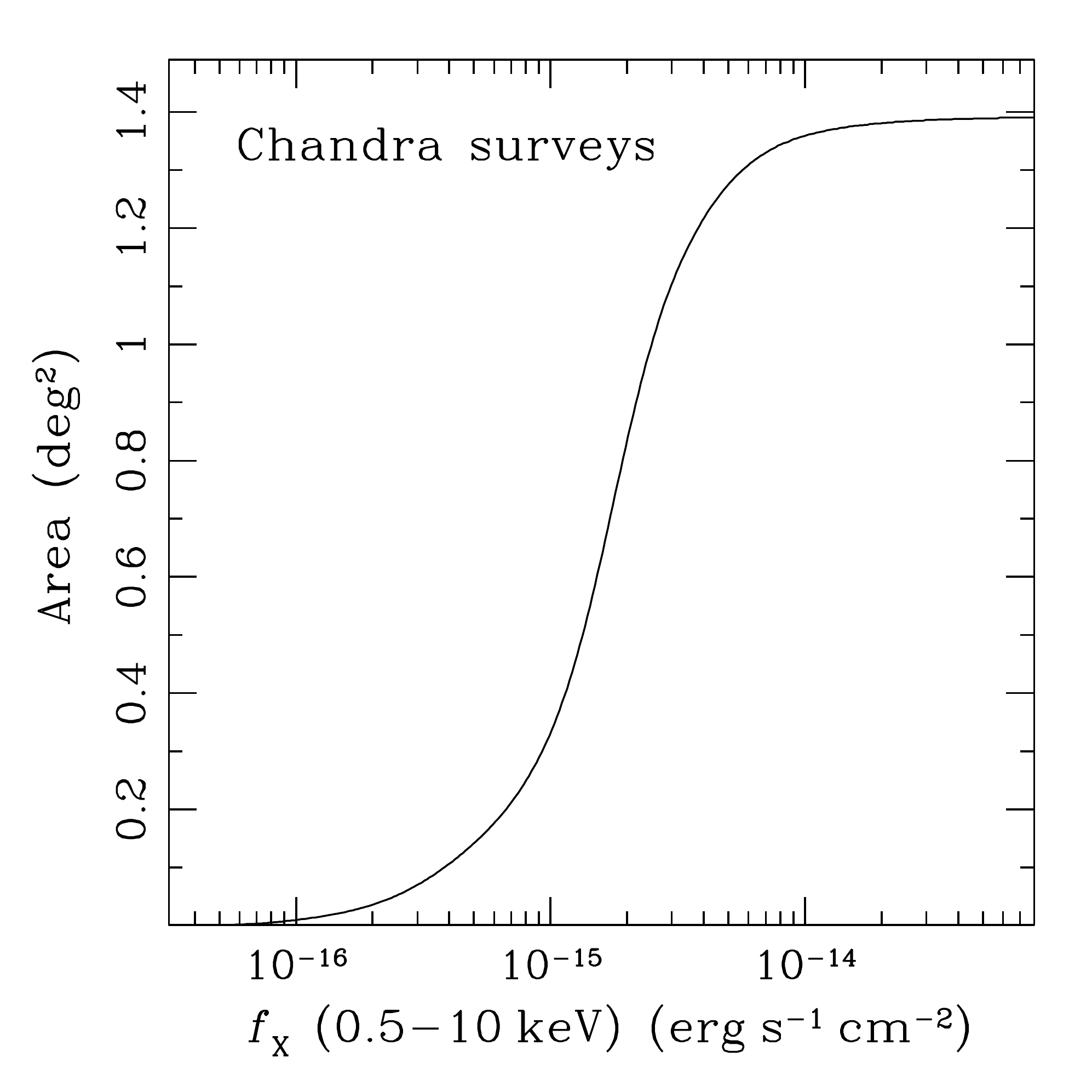}
\includegraphics[height=0.85\columnwidth]{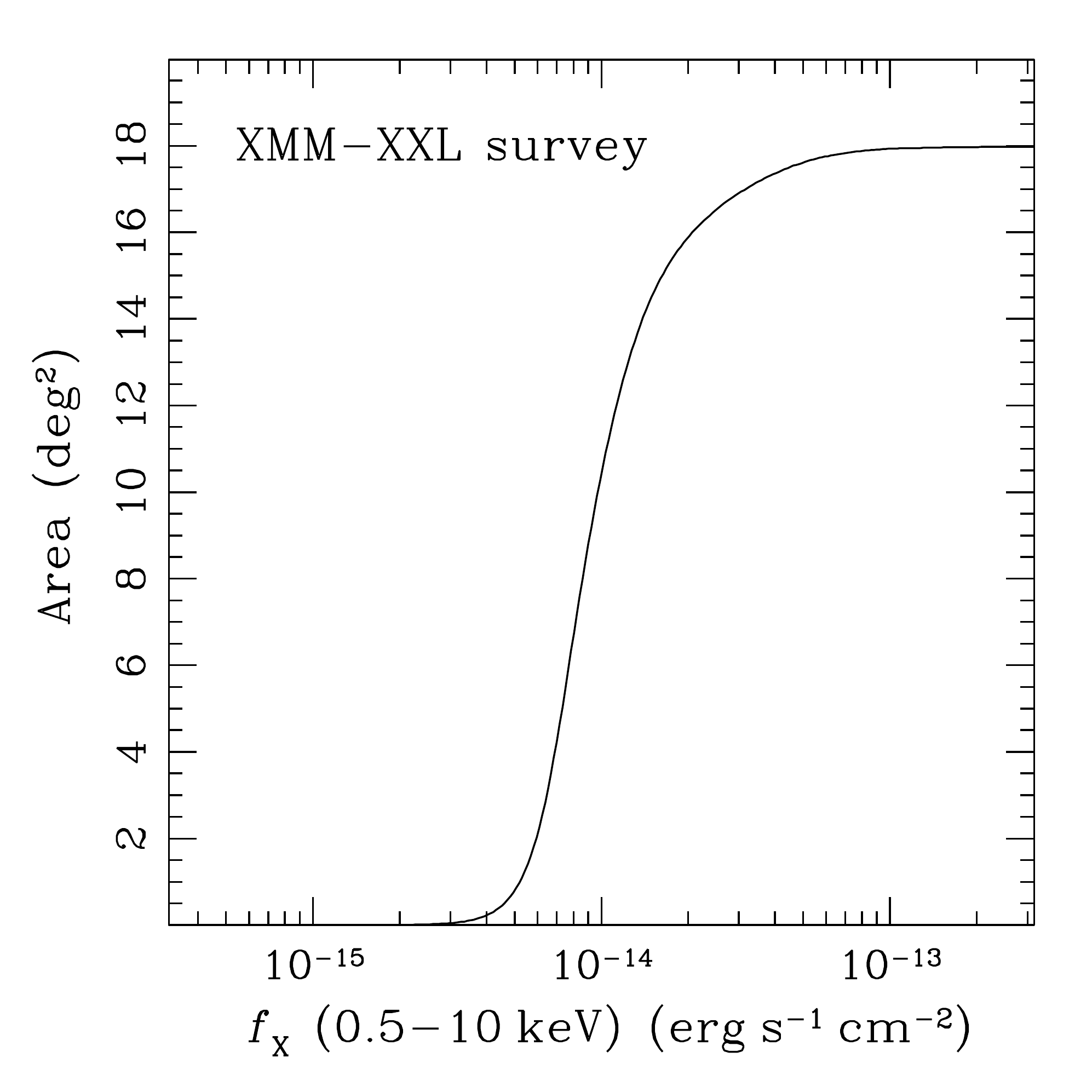}
\end{center}
\caption{X-ray Sensitivity curves for the combined Chandra surveys
  used in the analysis (left panel) and the XMM-XXL survey (right
  panel). 
}\label{fig_sense}
\end{figure*}

\subsection{Chandra survey fields}\label{sec_data_chandra}

The Chandra observations of the  CDFS, CDFN, AEGIS, ECDFS and C-COSMOS
were  analysed in  a homogeneous  way  by applying  the reduction  and
source detection methodology  described by \cite{Laird2009}.  Specific
details on  the analysis of the  4\,Ms CDFS and the  2\,Ms CDFN fields
are presented  by \cite{Rangel2013}.  The Chandra survey  of the AEGIS
field has two tiles. The wide  and shallow one (AEGIS-W) consists of 8
Chandra  pointings  of 200\,ks  each.   These  data  are described  by
\cite{Laird2009}.   The  deep survey  of  the  AEGIS field  (AEGIS-XD)
increased  to a  total of  800\,ks the  exposure time  of  the central
regions of the AEGIS-W.  The  additional data overlap with the central
3 of the original 8 Chandra pointings observed as part of the AEGIS-W.
The AEGIS-XD survey data reduction and source catalogue generation are
described by \cite{Nandra2015}.

The X-ray  sources used in this  paper are detected  in the 0.5-7\,keV
(full)   spectral  band   with  Poisson   false   detection  threshold
$<4\times10^{-6}$.   The count rates  in the 0.5-7\,keV  band are
converted  to fluxes  in  the 0.5-10\,keV  band  assuming a  power-law
spectral  index   with  $\Gamma=1.9$.    This  is  steeper   than  the
$\Gamma=1.4$ adopted  for the X-ray  flux estimation in  the published
catalogues  of  the  CDFS,  CDFN and  AEGIS  fields  \citep{Laird2009,
Rangel2013, Nandra2015}.  The choice  of the $\Gamma=1.9$ is motivated
by  the  fact  that  at  high redshift,  $z\ga3$,  the  observer-frame
0.5-7\,keV band  corresponds to harder rest-frame  energies, which are
least affected by obscuration.  A  diagnostic of the spectral shape of
X-ray sources is their hardness ratio defined as HR=(H-S)/(H+S), where
S, H are  the observed count rates in the  0.5-2 and 2-7\,keV spectral
bands, respectively. For sources in the range $z=3-5$ we find a median
hardness  ratio $\rm  HR  \approx  -0.2$. This  is  consistent with  a
power-law X-ray  spectrum with $\Gamma \approx 1.8$,  i.e.  similar to
the   mean    spectral   index   of   the    intrinsic   AGN   spectra
\citep[$\Gamma\approx1.9$, ][]{Nandra1994}. We therefore choose to fix
$\Gamma=1.9$ for  the determination of  fluxes. We note  however, that
the  choice of  $\Gamma$  (1.4 vs 1.9) has a  small  impact  on the
results presented in this paper. 

Sensitivity  curves,  which measure  the  total  survey  area that  is
sensitive  to sources of  a particular  flux are  calculated following
methods   described  in  \cite{Georgakakis2008_sense}.    The  overlap
between the ECDFS  and the 4\,Ms CDFS or the  AEGIS-W and the AEGIS-XD
is  accounted for  by defining  independent spatial  regions  for each
survey.  Spatial  masks  that  describe  both the  boundaries  of  the
optical/infrared imaging of each  field and regions of poor photometry
because of  bright stars \citep{Aird2015} are also  taken into account
in the X-ray sensitivity  calculations.  The sensitivity curves in the
0.5-10\,keV          band           are          presented          in
Figure~\ref{fig_sense}. Table~\ref{table_data}  presents the number of
X-ray  sources in each  field.  The  same spatial  masks used  for the
construction of sensitivity  maps are used to filter  the X-ray source
catalogue.

The optical identification of the X-ray sources in the CDFN, AEGIS-XD.
AEGIS-W, ECDFS and  C-COSMOS fields are based on  the Likelihood Ratio
method    \citep{Sutherland_and_Saunders1992}   as    implemented   in
\cite{Aird2015}. The  multiwavelength associations of  the 4\,Ms CDFS
X-ray sources  are presented by \cite{Hsu2014}. They  apply a Bayesian
methodology,  based  on  the  work of  \cite{Budavari_Szalay2008},  to
different   catalogues   available  in   that   field  including   the
CANDELS/$H$-band selected photometry  presented by \cite{Guo2013}, the
Taiwan ECDFS Near-Infrared  Survey \citep[TENIS;][]{Hsieh2012} and the
MUSYC/$BVR$-selected  catalog of  \cite{Cardamone2010}. The  number of
X-ray  sources with secure  optical or  infrared counterparts  in each
field are presented in Table~\ref{table_data}.

Extensive spectroscopic campaigns have  been carried out in the fields
of choice. For  the CDFN, ECDFS and AEGIS-W we  use the compilation of
spectroscopic  redshifts presented by  \cite{Aird2015}. In  the 4\,Ms
CDFS we  use the  spectroscopic redshifts compiled  by \cite{Hsu2014}.
In the  case of AEGIS-XD  we use the spectroscopic  redshift catalogue
presented by  \citep{Nandra2015}.  Redshifts in the  C-COSMOS are from
the   public   releases    of   the   VIMOS/zCOSMOS   bright   project
\citep{Lilly2009}   and  the   Magellan/IMACS   observation  campaigns
\citep{Trump2009}, as  well as the compilation of  redshifts for X-ray
sources  presented  by   \cite{Civano2012}.     The  spectroscopic
redshifts  used in  this paper  have  quality flags  in the  published
catalogues from which they  were retrieved that indicate a probability
better than $\approx95$\% of being correct.

\begin{figure}
\begin{center}
\includegraphics[height=0.85\columnwidth]{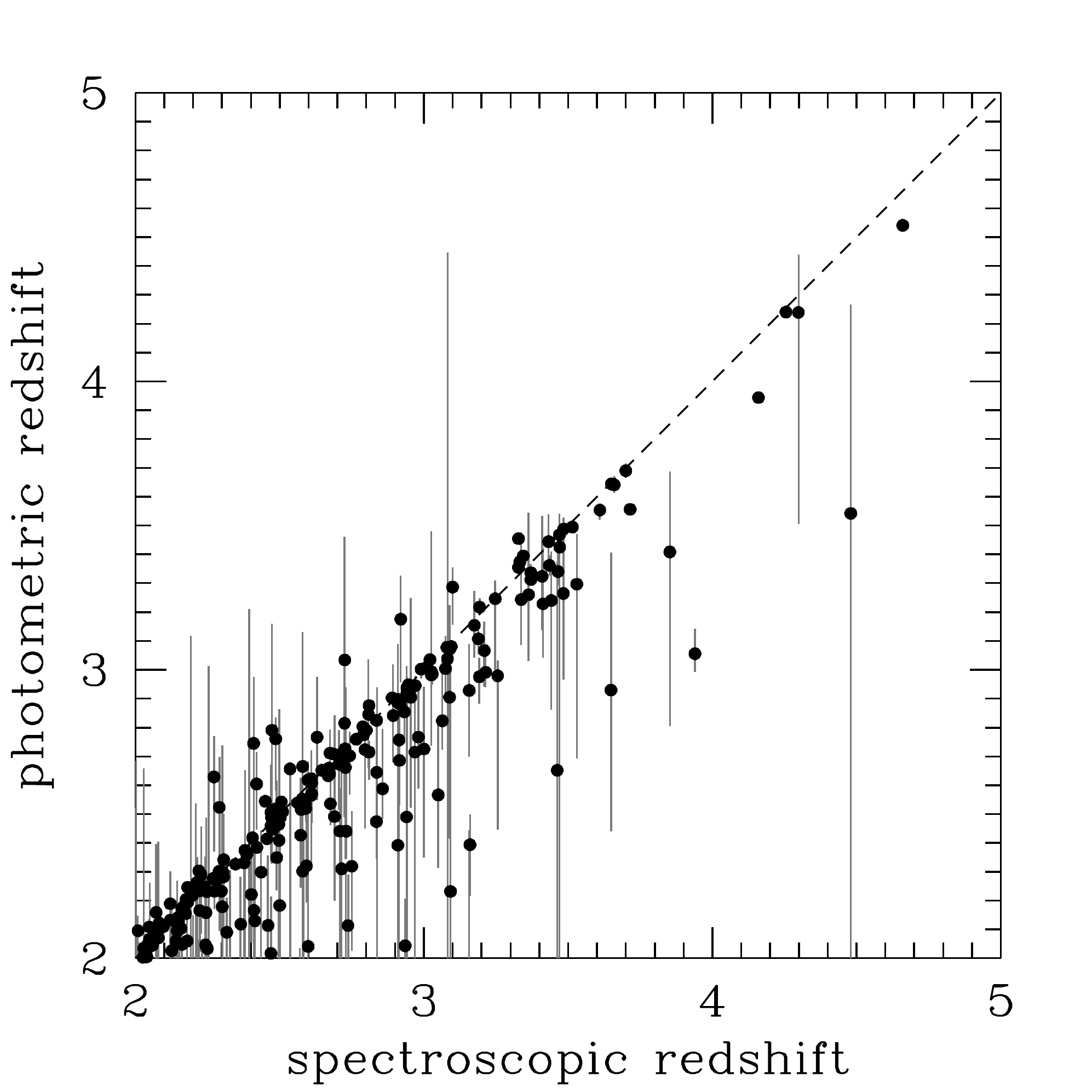}
\end{center}
\caption{Spectrocopic vs photometric redshift measurements for the
  sample of X-ray selected sources used in this paper. The data points
  correspond to the median  value of the PDZ. The errorbars correspond
  to the 90\% confidence interval around the median. 
}\label{fig_zz}
\end{figure}

\begin{figure}
\begin{center}
\includegraphics[height=0.85\columnwidth]{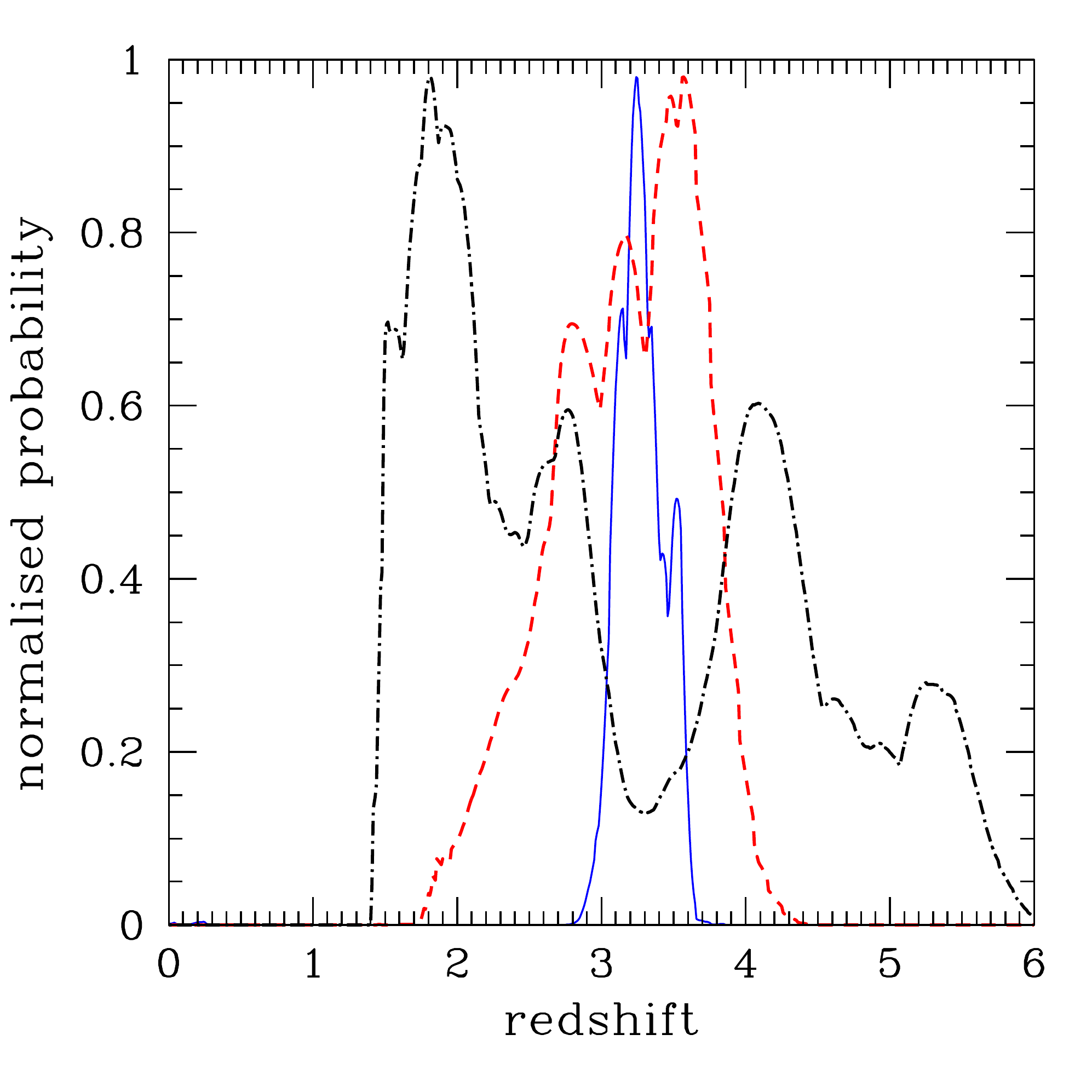}
\end{center}
\caption{Examples  of  photometric  redshift Probability  Distribution
Functions for  X-ray selected AGN  in the AEGIS field.   The different
colours correspond  to different  sources. There is  a variety  of PDZ
shapes in the sample,  including uni-modal and relatively narrow (blue
solid),  unimodal  but  broad  (red  dashed)  and  multi-modal  (black
dot-dashed).}\label{fig_pdz}
\end{figure}

For  X-ray sources  without  spectroscopy,  photometric redshifts  are
estimated using  the multiwavelength photometric  catalogues available
for each survey field. The  photometric redshifts of the X-ray sources
in the 4\,Ms  CDFS, the AEGIS-XD and the COSMOS  fields are determined
following    the   methodology    described   by    \cite{Salvato2009,
  Salvato2011}.  Specific details  can be found in Hsu  et al.  (2014;
4\,Ms  CDFS), Nandra  et  al.   (2015; AEGIS-XD)  and  Salvato et  al.
(2011;  COSMOS field).   The estimated  rms scatter  of the  X-ray AGN
photometric redshifts  is $\sigma_{\Delta z/(1  + z)} =  0.016$, 0.014
and  0.04  for   the  C-COSMOS,  4\,Ms  CDFS   and  AEGIS-XD  samples,
respectively.  The corresponding outlier  fraction, defined as $\Delta
z/(1 + z) > 0.15$, is about 5-6\% in all three fields.  In the case of
ECDFS, AEGIS-W and CDFN we  use the photometric redshifts estimated by
\cite{Aird2015}. For these  fields $\sigma_{\Delta z/(1 +  z)} = 0.06$
and the  outlier fraction is about  15\%.  The latter value  is larger
than in the C-COSMOS, 4\,Ms CDFS  and AEGIS-XD fields. This is related
to differences in the  methodology of estimating photometric redshifts
and ultimately to the choice of template SEDs used in the calculation.
Nevertheless, the  photometric redshifts estimated  by \cite{Aird2015}
are assigned  appropriately larger uncertainties, approximated  by the
corresponding Probability  Distribution Functions (PDZ),  that reflect
the higher outlier fraction.   Figure \ref{fig_zz} plots spectroscopic
vs  photometric  redshifts for  the  sample  used  in this  paper  and
illustrates  the   overall  quality   of  the   photometric  redshifts
estimates.  In the  X-ray luminosity function calculations  we use the
full  photometric redshift  PDZ.These  are typically  unimodal but  at
increasing redshift they broaden and  secondary peaks may also appear.
Also the \cite{Aird2015} PDZs are  typically broader than those in the
C-COSMOS, 4\,Ms  CDFS and AEGIS-XD  fields.  Examples of PDZs  used in
this paper are  shown in Figure~\ref{fig_pdz}.  Table~\ref{table_data}
presents the number of photometric and spectroscopic redshifts in each
field, both  total and  in the intervals  $z=3-4$ and  $4-5$.  Sources
without optical identifications and hence, without redshift estimates,
are a  minority in the  Chandra surveys  sample, 107 in  total.  These
sources can be  either moderate redshift ($z\approx1-3$)  AGN with red
SEDs  because  of e.g.   obscuration  and/or  old stellar  populations
\citep{Koekemoer2004,  Schaerer2007, Rodighiero2007,  DelMoro2009}, or
high redshift  systems ($z\ga3$).  Since the  redshift distribution of
these  sources is  not  known they  are  assigned a  flat  PDZ in  the
redshift  interval $z=1-6$  and zero  at other  redshifts. Fixing  the
lower limit of  the redshift range above to a  value between $z=0$ and
$z=2$ does not change the results and conclusions.

 Additionally we have tested  that the differences in the accuracy
and  outlier fraction of  the photometric  redshifts in  the different
fields used  in this work  do not affect  the final results.   For the
COSMOS,  AEGIS-XD and  4\,Ms CDFS  we can  substitute  the photometric
redshift PDZs adopted  in this paper (based on  the methods of Salvato
et al.  2009,  2011) with those estimated following  the methodoloy of
Aird et  al. (2015).  For these  fields we can  therefore estimate the
X-ray  luminosity function at  $z=3-5$ (see  next sections)  using two
different sets of photometric redshifts,  those of Aird et al.  (2015)
and those determined following Salvato et al. (2009, 2011). We find no
systematic  differences  between  the  X-ray luminosity  functions  at
$z=3-5$  estimated  using  the  two independent  photometric  redshift
catalogues.

\subsection{XMM-XXL survey data}\label{sec_data_xxl}

The  Chandra  survey fields  are  complemented  by  the wide-area  and
shallow XMM-XXL  survey (PI: Pierre).  The  XMM-Newton observations of
this programme  cover a  total of about  $\rm 50\,deg^2$ of  sky split
into  two  nearly  equal  area  fields.   In this  paper  we  use  the
equatorial   region  of   the   XMM-XXL,  which   overlaps  with   the
Canada-France-Hawaii  Legacy Survey  (CFHTLS) W1  field and  covers an
area of  about $\rm  25\,deg^2$.  The  approximate centre  of this
field  lies at  right ascension  $\alpha=10:22:42.20$  and declination
$\delta=-04:52:51.03$.

The reduction of the XMM data, the construction of the X-ray catalogue
and the association of the X-ray sources with optical counterparts are
described  by  \cite{Liu2015}  based   on  the  methods  presented  in
\cite{Georgakakis_Nandra2011}. In  brief, the X-ray  data reduction is
carried out  using the XMM  Science Analysis System (SAS)  version 12.
We analyse  XMM-Newton observations  related to the  XMM-XXL programme
that were made public prior  to 23 January 2012. XMM-XXL data observed
after that  date are not  included in the  analysis.  As a  result our
final  catalogue of  the equatorial  XMM-XXL field  misses  about $\rm
5\deg^2$ worth of X-ray coverage.  The {\sc epchain} and {\sc emchain}
tasks of  {\sc sas} are employed  to produce event files  for the EPIC
\citep[European Photon  Imaging Camera;][]{Struder2001, Turner2001} PN
and MOS detectors respectively.  Flaring periods resulting in elevated
EPIC  background  are  identified  and excluded  using  a  methodology
similar  to  that described  by  \cite{Nandra2007_Fe}.   We use  X-ray
sources detected in the  $0.5-8$\,keV spectral band with Poisson false
detection probability of $<4\times10^{-6}$.  The final sample consists
of  7493 unique  sources detected  in 0.5-8\,keV  spectral  band.  The
fluxes listed  in the  final source catalogue  are in  the 0.5-10\,keV
band   assuming  a   power-law  spectral   energy   distribution  with
$\Gamma=1.4$.   These  X-ray  sources  are  matched  to  the  SDSS-DR8
photometric catalogue  \citep{Aihara2011} using the Maximum-Likelihood
method (Sutherland \& Saunders  1992).  We assign counterparts to 3798
sources with Likelihood Ratio $\rm  LR>1.5$.  At that cut the spurious
identification rate  is about 6\%  and the total number  of 0.5-8\,keV
detected  sources  with  optical   counterparts  is  3798  (see
Table~\ref{table_data}). 

Redshifts  for the XMM-XXL  X-ray sources  are from  several follow-up
spectroscopic campaigns.  The XMM-XXL field overlaps with the SDSS-III
Baryon  Oscillation  Spectroscopic Survey  \citep[BOSS;][]{Dawson2013}
programme,  which  provides  spectroscopy  for  UV/optically  selected
broad-line   QSOs  and   luminous  red   galaxies.   \cite{Stalin2010}
presented  spectroscopy for  X-ray  sources selected  in the  original
XMM-LSS  survey \citep{Clerc2014},  which  is part  of the  equatorial
XMM-XXL survey field.  Most of the redshifts however, are from a total
of  five special SDSS  plates dedicated  to follow-up  spectroscopy of
X-ray sources as part of  the Ancillary Programs of SDSS-III.  The
overlap between  those plates  and the XMM-XXL  survey region  is $\rm
17.98\,deg^{-2}$. Targets were selected to have $f_X(\rm 0.5-10\,keV,
\Gamma=1.4) > 10^{-14}  \, erg \, s^{-1} \,  cm^{-2}$ and $15<r<22.5$,
where $r$  is either the SDSS  PSF magnitude in the  case of optically
unresolved  sources (SDSS  type=6)  or the  SDSS  model magnitude  for
resolved  sources.     Specific  details  on  these  spectroscopic
observations, including spectral classification, visual inspection and
redshift quality assessment  are presented by \cite{Menzel2015}.  The
total number of XMM-XXL sources with secure spectroscopic measurements
are presented in Table~\ref{table_data}.  Also shown in this table are
the  number of sources  with spectroscopic  redshifts in  the interval
$z=3-5$.

Although when constructing  the X-ray source catalogue  of the XMM-XXL
field  fluxes  are  estimated  for a  power-law  X-ray  spectrum  with
spectral  index $\Gamma=1.4$,  in the  rest of  the analysis  we adopt
$\Gamma=1.9$  for   the  calculation   of  fluxes,   luminosities  and
sensitivity maps.  This is because at  the depth of the XMM-XXL survey
AGN  at $z>3$  are  powerful QSOs  with  $L_X(\rm 2  -  10\, keV)  \ga
10^{44}\,  erg \,s^{-1}$.   The fraction  of obscured  AGN among  such
luminous   sources   is   a    decreasing   function   of   luminosity
\citep{Ueda2003, Akylas2006,  Merloni2014, Ueda2014,  Buchner2015} and
therefore  a  spectral index  of  $\Gamma=1.9$,  which represents  the
intrinsic unobscured  power-law X-ray spectrum of  both local Seyferts
\citep[e.g][]{Nandra1994}    and     luminous    high-redshift    QSOs
\citep[e.g.][]{Vignali2005, Shemmer2005, Just2007}  is appropriate for
this population.   The 0.5-10\,keV  fluxes estimated  for $\Gamma=1.9$
are  about 35\%  fainter  than those  for  $\Gamma=1.4$.  The  XMM-XXL
sensitivity   curve   in   the    0.5-10\,keV   band   is   shown   in
Figure~\ref{fig_sense} and is estimated following methods described in
\cite{Georgakakis_Nandra2011}. In  the calculation of  the sensitivity
curve  we only  consider  the overlapping  area  between the  SDSS-III
Ancillary Programs  spectroscopic plates used to  target X-ray sources
and the  XMM-XXL survey  region. We  also take  into account  the flux
limit for  follow-up spectroscopy $f_X(\rm 0.5-10\,keV,  \Gamma=1.4) >
10^{-14} \, erg \, s^{-1} \,  cm^{-2}$. This limit appears as a smooth
drop in area rather than a sharp cut in Figure~\ref{fig_sense} because
the Poisson probability  of measuring a flux above this  limit is used
to determine the sensitivity curve.

\begin{figure}
\begin{center}
\includegraphics[height=0.9\columnwidth]{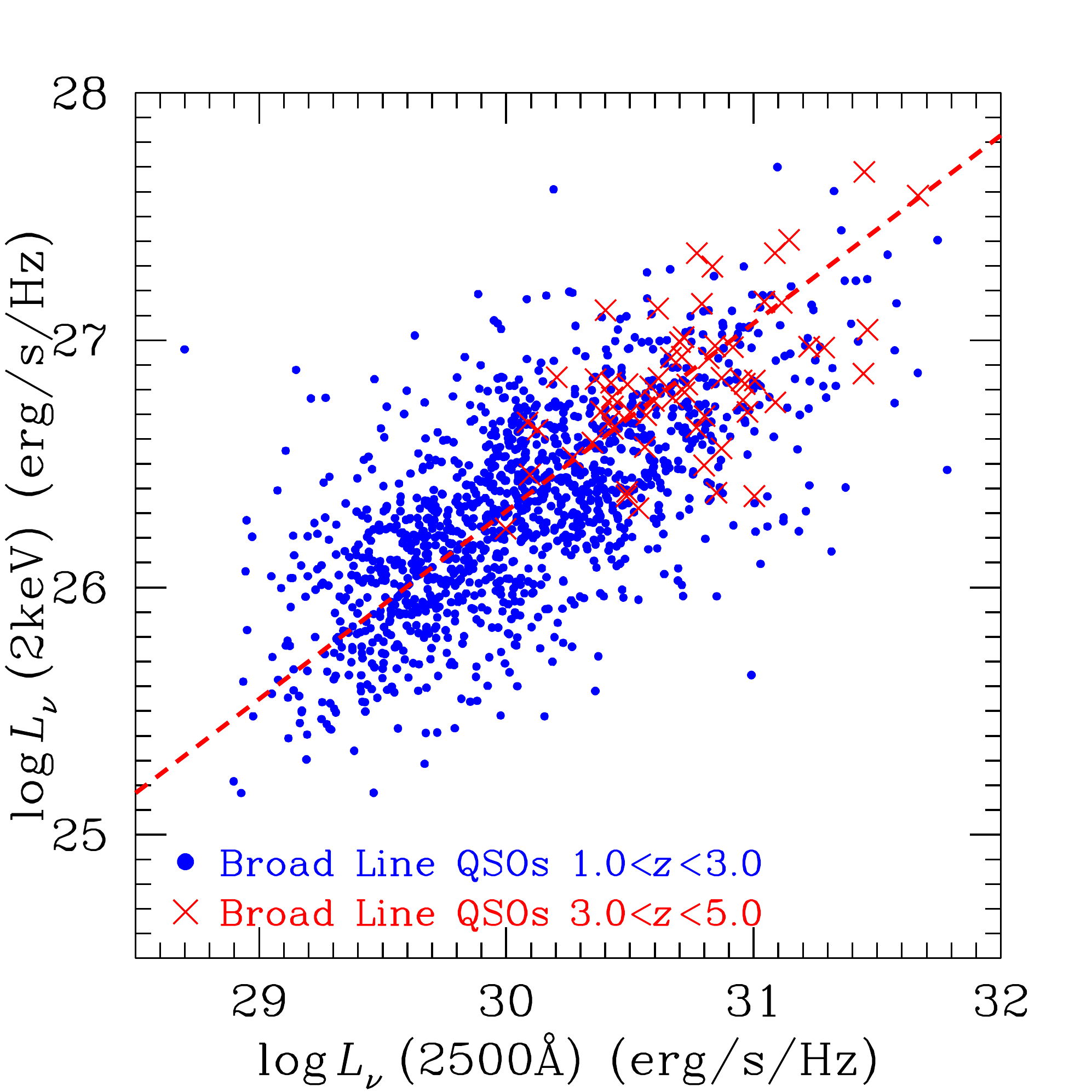}
\end{center}
\caption{Monochromatic  2\,keV X-ray luminosity,  $L_\nu(\rm 2\,keV)$,
plotted  as a  function  of monochromatic  2500\,\AA\, UV  luminosity,
$L_\nu(\rm 2500\,\AA)$.   The data points (cross or  dots) are XMM-XXL
X-ray selected broad line  QSOs with secure spectroscopic redshifts in
the interval $z=1-5$. Sources  at $z>3$ are highlighted with different
symbols  (crosses).  The  red  dashed line  is  the bisector  best-fit
$L_\nu(\rm 2\,keV)-L_\nu(\rm 2500\,\AA)$  relation determined by Lusso
et  al.  [2010;  $\log  L_{\rm  2\,keV}  =  0.760  \;  \log  L_\nu(\rm
2500\,\AA)  + 3.508$].   For  the XMM-XXL  QSOs  the X-ray  luminosity
density at  2\,keV is  estimated from the  0.5-2\,keV flux  assuming a
power-law   X-ray  spectrum   with  index   $\Gamma=1.9$.      The
2500\,\AA\,  monochromatic  luminosity  is  determined from  the  SDSS
photometry.   For  the  k-corrections   we  adopt  the  simulated  QSO
templates of \protect\cite{McGreer2013}.  For  a QSO with redshift $z$
the SDSS photometric bands with effective wavelengths that bracket the
wavelength  $2500\times(1+z)\rm  \,\AA\,$  are identified.   The  mean
model QSO  SED at that  redshift is then  scaled to the  observed SDSS
optical magnitudes  in those two bands.   The monochromatic luminosity
at  2500\,\AA\, is  then  estimated  from the  scaled  model SED.   At
redshifts  $z\ga2.7$  the   rest-frame  2500\,\AA\,  lies  beyond  the
effective  wavelength of  the  SDSS $z$-band  (9134\,\AA).  For  these
sources  the  flux  density  at  $2500\times(1+z)\rm  \,\AA\,$  is  an
extrapolation using  the model  SED. The results  do not change  if we
simply linearly interpolate between the observed flux densities of the
SDSS   bands   that    bracket   the   rest-frame   2500\,\AA.    This
model-independent  approach  however,  does  not  allow  extrapolation
beyond $z\ga2.7$.  }\label{fig_l2l2500}
\end{figure}

\begin{figure}
\begin{center}
\includegraphics[height=0.9\columnwidth]{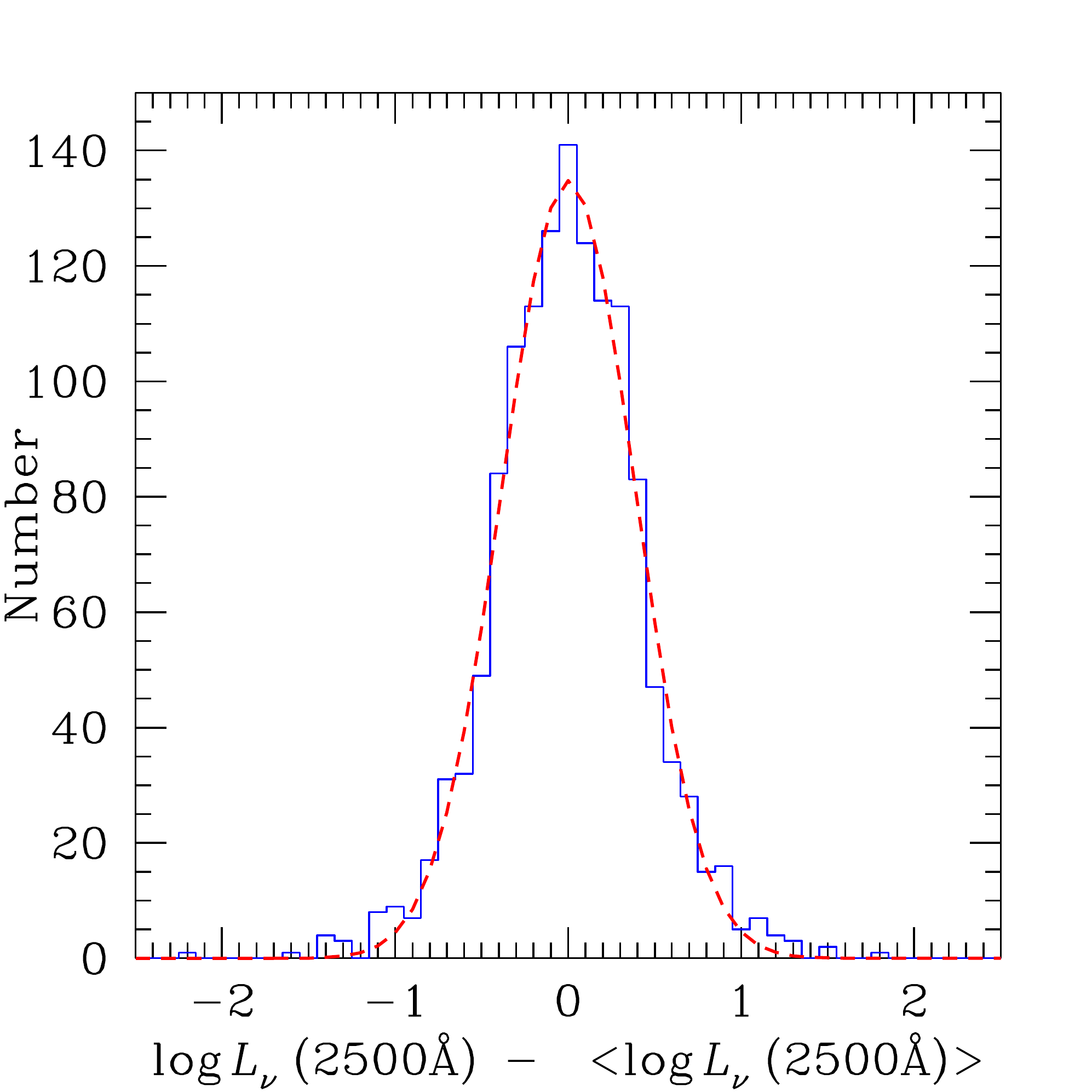}
\end{center}
\caption{Histogram   of   the   difference  $\Delta   \log   L_\nu(\rm
  2500\,\AA)$  between  the  monochromatic 2500\,\AA\,  luminosity  of
  broad-line    QSOs    in    the     XMM-XXL    (blue    points    of
  Figure~\ref{fig_l2l2500}) and the mean $<\log L_\nu(\rm 2500\,\AA)>$
  value predicted  by the bisector  best-fit relation of Lusso  et al.
  (2010;  red dashed  line  of Figure~\ref{fig_l2l2500})  for a  given
  $\log L_\nu(\rm  2\,keV)$.  The red  dashed line shows  the best-fit
  Gaussian  distribution.   The mean  value  of  that distribution  is
  consistent  with zero  and the  standard deviation  is $\sigma=0.4$.}
\label{fig_hist_l2500}
\end{figure}

\begin{figure}
\begin{center}
\includegraphics[height=0.9\columnwidth]{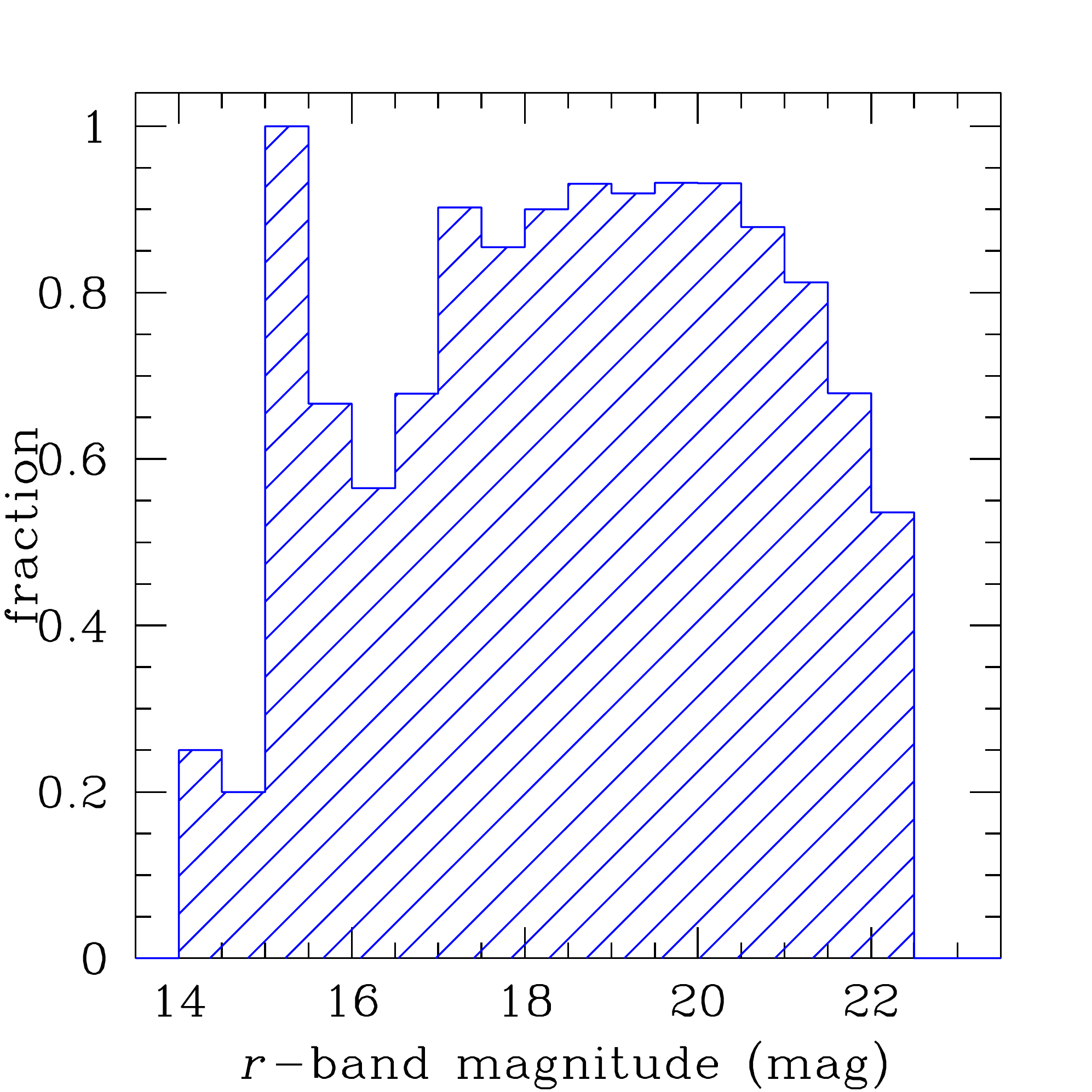}
\end{center}
\caption{Fraction  of XMM-XXL X-ray  sources with  successful redshift
measurement as  a function  of the $r$-band  magnitude of  the optical
counterpart.  The fraction is defined  as the ratio between the number
of  potential targets  [$f_X(\rm 0.5-10\,keV,  \Gamma=1.4)>10^{-14} \,
erg  \,  s^{-1}  \,  cm^{-2}$  and secure  optical  counterparts  with
$r<22.5$\,mag] and X-ray sources with successful redshift measurements
and fluxes/magnitudes within the above cuts.  }\label{fig_zeff_hist}
\end{figure}

\begin{figure}
\begin{center}
\includegraphics[height=0.9\columnwidth]{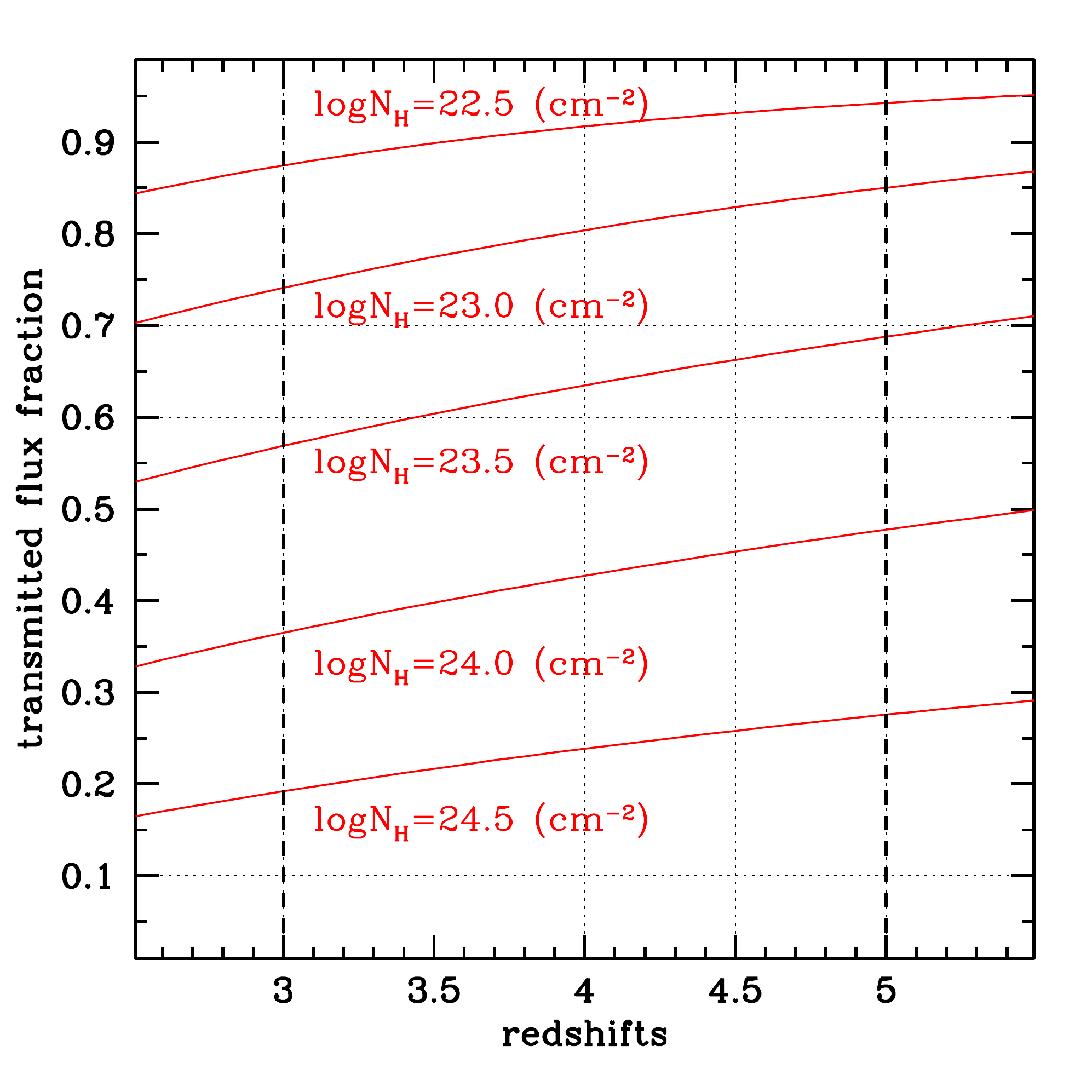}
\end{center}
\caption{Impact of different levels  of obscuration on the 0.5-10\,keV
flux of a  source as a function of redshift. The  vertical axis is the
ratio of the flux of an  AGN obscured by column density $N_H$ relative
to the flux of the same  source in the case of zero obscuratrion.  The
red curves correspond to different  levels of obscuration in the range
$\log N_H =  22.5 - 24.5$ ($\rm cm^{-2}$  units).  The vertical dashed
lines  mark the  redshift  interval of  interest,  $z=3-5$. For  fixed
obscuration and  intrinsic luminosity the flux of  higher redshift AGN
is less affected.  The differential flux suppression between $z=3$ and
$z=5$ is small, $\la10$\%.  For the calculation of X-ray k-corrections
we       adopt      the       model       X-ray      spectral       of
\protect\cite{Brightman2011_torus}.   These are  based on  Monte Carlo
simulations of an  illuminating source at the centre  of a sphere with
constant  density and  a conical  region (apex  at the  centre  of the
sphere)  cut-off  to  approximate  a  toroidal  geometry.    These
simulations   take   into   account   both  Compton   scattering   and
photoelectric  absorption  of  the  X-ray  photons  by  the  obscuring
medium.   We adopt $\Gamma=1.9$  for the  intrinsic AGN  spectrum, an
opening angle of  the conical region of $60$\,deg  and a viewing angle
of  $\rm 45\,deg$,  i.e.  a  line-of-sight intersecting  the obscuring
material.  }\label{fig_suppression}
\end{figure}

\section{Methodology}

\subsection{X-ray Luminosity function estimation: Chandra survey fields}

A  Bayesian approach  is adopted  for the  determination of  the X-ray
luminosity  function.  The  X-ray  sources detected  in  a survey  are
essentially Poisson realisations of a  parent sample and therefore the
likelihood can be written as  the product of the Poisson probabilities
of individual  sources.  Following  the works  of \cite{Marshall1983},
\cite{Loredo2004},   \cite{Aird2010}    and   \cite{Buchner2015}   the
likelihood can be written as

\begin{equation}\label{eq_likeli}
\begin{split}
 \mathcal{L}( d_i \;|\; \theta) & =  e^{-\lambda} \times \\
                                & \prod_{i=1}^{N} \int {\rm d} \log L_\mathrm{X} \, \frac{{\rm d}
V}{{\rm  d}   z}  {\rm  d}  z   \;  p(  d_i   |\;  L_\mathrm{X},z)  \;
\phi(L_\mathrm{X},z \;|\; \mathbf{\theta}),
\end{split}
\end{equation}

\noindent where ${\rm  d}V/{\rm d}z$ is the comoving  volume per solid
angle  at  redshift $z$,  $d_i$  signifies  the  dataset and  $\theta$
represents   the  parameters   of  the   luminosity   function  model,
$\phi(L_\mathrm{X},z   \;|\;  \mathbf{\theta})$,   that   are  to   be
estimated.   The multiplication  is  over all  sources,  $N$, and  the
integration is  over redshift and  X-ray luminosity. The  quantity $p(
d_i  |\; L_\mathrm{X},z)$ is  the probability  of a  particular source
having redshift  $z$ and X-ray luminosity $L_{\rm  X}$.  This captures
uncertainties   in   the  determination   of   both  redshifts   (e.g.
photometric  redshifts  measurements)  and  X-ray  fluxes  because  of
Poisson     statistics     and      the     Eddington     bias.     In
equation~\ref{eq_likeli}, $\lambda$ is the expected number of detected
sources in a survey for a particular set of model parameters $\theta$

\begin{equation}\label{eq_lambda}
%\begin{split}
 \lambda =  \int {\rm d} \log L_\mathrm{X} \, \frac{{\rm d} V}{{\rm  d}
   z}  {\rm  d}  z   \;  A(L_\mathrm{X},z)  \; \phi(L_\mathrm{X},z \;|\; \mathbf{\theta}).
%\end{split}
\end{equation}

\noindent where,  $A(L_\mathrm{X}, z)$  is the sensitivity  curve that
quantifies the survey  area over which a source  with X-ray luminosity
$L_X$ and redshift  $z$ (and hence flux $f_X$)  can be detected.  Note
that the selection function term,  $A(L_X, z)$, is not included within
the  integral of  equation 1.   The  reader is  referred to an extensive
discussion in Loredo (2004) on that point.

The  goal of  this paper  is the  estimation of  the  X-ray luminosity
function in the redshift  interval $z=3-5$. However, defining a sample
of  X-ray   AGN  in  a   relatively  narrow  redshift  range   is  not
straightforward.  This  is because the  redshifts of many  sources are
determined  by photometric methods  and therefore  have uncertainties,
which  are  not  negligible  compared  to the  size  of  the  redshift
interval.   It  may  happen  for  example,  that  the  errors  of  the
photometric  redshift of a  particular source  straddle one  (or both)
boundaries of the redshift range of interest (see Fig.~\ref{fig_pdz}).

We deal with this difficulty by  simply using all sources in the X-ray
sample and splitting the luminosity function model into two terms

\begin{equation}\label{eq_split_xlf}
\begin{split}
\phi(L_\mathrm{X},z \;|\; \mathbf{\theta})  = 
 \phi_1(L_\mathrm{X},z \in [z_1, z_2] \;|\; \theta_1) & + \\
\phi_2(L_\mathrm{X}, z \notin [z_1,z_2] \;|\; \theta_2).
\end{split}
\end{equation}

\noindent The  first term $\phi_1$  refers to the  luminosity function
within the redshift interval of interest, $z=z_1-z_2$, and has its own
set of parameters, $\theta_1$.  The second term, $\phi_2$, corresponds
to the X-ray  AGN space density outside that redshift  range and has a
different set of parameters, $\theta_2$, which are treated as nuisance
parameters.   The advantage  of this  approach  is that  it allows  an
estimate of the AGN space density at $z=3-5$ nearly independent of the
shape and  form of the evolution  of the X-ray luminosity  function at
lower  redshift. In  any X-ray  selected sample  the bulk  of the  AGN
population lies at low redshift $z<3$.  This may introduce systematics
in the determination of the AGN  space density at $z\ga3$, if the same
evolutionary law  is fit to  the data  across all redshifts.   In this
case it is possible that the  determination of the model parameters is
dominated by  the regions  of the redshift/luminosity  parameter space
with  the  most data.   Our  approach  minimises  the impact  of  this
potential  source of  bias.   In  our analysis  the  $\phi_2$ term  is
modelled as a step function, i.e. the sum of constants that correspond
to the AGN space densities  at different luminosity and redshift bins.
The  values  of  these  constants  are  determined  by  the  data  via
equation~\ref{eq_likeli}.       For       the      calculation      of
equation~\ref{eq_likeli} assumptions  need to be made  on the redshift
errors.  For sources with photometric redshift determinations we adopt
the corresponding redshift Probability  Distribution Function (PDZ) as
a measure  of the uncertainty.    Spectroscopic redshifts  in the
  sample   have   reliabilities    $\ga95\%$   and   therefore   their
  corresponding  PDZs  are  assumed  to  be  delta  functions  at  the
  spectroscopic redshift of the source. Sources without optical
counterparts are assigned a flat PDZ in the redshift range $z=1-6$
(see Section \ref{sec_data_chandra}).

Poisson statistics are used to determine the flux distribution that is
consistent  with  the  extracted  source  and  background  counts.   A
power-law  X-ray  spectrum  with   $\Gamma=1.9$  is  adopted  in  this
calculation.  The  flux distribution in  the 0.5-10\,keV band  is then
convolved with the PDZ to estimate the luminosity distribution of each
source at  rest-frame energies 2-10\,keV.  The  relevant k-corrections
also  assume  a  power-law  X-ray  spectrum  with  $\Gamma=1.9$.   The
two-dimensional probability distribution in $L_X$  and $z$ is the term
$p(  d_i |\;  L_\mathrm{X},z)$ of  equation~\ref{eq_likeli}.    In
  practice we  use importance sampling \citep{NR1992}  to evaluate the
  integral of equation~\ref{eq_likeli}.  For each source we draw $L_X$
  and  $z$  samples  based  on   the  PDZ  and  Poisson  X-ray  counts
  distribution  of  that  source.   The luminosity  function  is  then
  evaluated  for  each  sample  point,  $L_X,  z$.   The  integral  of
  equation~\ref{eq_likeli} is  simply the average  luminosity function
  of the sample.

\subsection{X-ray Luminosity function estimation: XMM-XXL field}

In the  case of  the XMM-XXL  field, only  spectroscopically confirmed
sources in  the redshift interval  $z=3-5$ are used.  For  that sample
there is no need to apply equation~\ref{eq_split_xlf} to determine the
corresponding  AGN  space  density.   The limitation  of  that  sample
however, is that it is both  X-ray and optical flux limited because of
the magnitude limit of the spectroscopic follow-up observations.  Both
cuts need to  be accounted for to infer the  X-ray luminosity function
via equation~\ref{eq_likeli}.  We do that  by exploiting the fact that
the $z>3$ XMM-XXL sample consists of  powerful [$L_X ( \rm 2-10 \, keV
  )\ga  10^{44}  erg   \,  s^{-1}$]  broad-line  QSOs.    We  use  the
well-established correlation between  monochromatic X-ray [$L_\nu (\rm
  2\,keV)$] and  UV [$L_\nu(\rm 2500\AA)$] luminosities  of broad-line
QSOs  \citep[e.g.][]{Steffen2006,  Just2007,  Lusso2010} to  link  the
observed optical magnitudes and X-ray fluxes of the sample and account
for the selection  effects.  Figure~\ref{fig_l2l2500} demonstrates the
correlation between $L_\nu (\rm 2\,keV)$ vs $L_\nu(\rm 2500\AA)$ using
X-ray selected broad-line QSOs from  the XMM-XXL field in the redshift
interval $z=1-5$.  The low  redshift cut  is to  avoid X-ray  AGN with
relatively low luminosities, for  which the UV/optical continuum shows
non-negligible  contribution  from  the  host galaxy.   We  adopt  the
\cite{Lusso2010} bisector best-fit $\log  L_\nu {(\rm 2\,keV)} = 0.760
\;  \log L_\nu  (\rm 2500\,\AA)  + 3.508$.   At a  given monochromatic
optical luminosity we  assume that the data points  scatter around the
above  relation  following  a   Gaussian  distribution  with  standard
deviation $\sigma$.  Figure~\ref{fig_hist_l2500} shows  that this is a
reasonable  assumption.  From  that figure  we estimate  $\sigma=0.4$.
Broad  Absorption  Line (BAL)  QSOs  represent  up  to about  26\%  of
optically    selected    samples   \citep{Hewett2003,    Reichard2003,
  Gibson2009},  and are  known to  be  X-ray faint  either because  of
absorption or  intrinsic X-ray  weakness \citep[e.g.][]{Gallagher2006,
  Luo2014}.  Such sources  do  not appear  to  skew the  distributions
plotted  in Figures~\ref{fig_l2l2500}~and~\ref{fig_hist_l2500}.   This
is  likely related  to  the bright  X-ray flux  limit  of the  XMM-XXL
survey, which  selects against  X-ray faint  Under assumptions  on the
optical      Spectral      Energy      Distribution      of      QSOs,
Figure~\ref{fig_hist_l2500} can be used  to estimate the SDSS $r$-band
optical magnitude distribution of X-ray sources of given $L_X$ and $z$
and then determine the fraction  of this distribution that is brighter
than the spectroscopic magnitude  limit of the survey, $r_{cut}=22.5$.
In this case the expected number of detected sources with the surveyed
area, $\lambda$, in equation~\ref{eq_likeli} can be rewritten

\begin{equation}\label{eq_xxl}
\begin{split}
 \lambda =  & \int {\rm d log} L_\mathrm{X} \, \frac{{\rm d} V}{{\rm d}  z}  {\rm  d}  z   \;  A(L_\mathrm{X},z) \;
 B(L_\mathrm{X},z, \;|\ r) \\
 & \times \phi(L_\mathrm{X},z \;|\; \mathbf{\theta}) \; \eta(r), 
\end{split}
\end{equation} 

\noindent where  $B (L_\mathrm{X},  z \;|\; r)$  is the  SDSS $r$-band
magnitude distribution of a source with $L_X$ and $z$.  The efficiency
factor $\eta(r)$ is the success rate of measuring secure redshifts for
X-ray  sources as  a  function  of the  SDSS  $r$-band magnitude.   It
accounts  for  the  fact  that  not  all  X-ray  sources  with  secure
counterparts      have      successful     spectroscopic      redshift
measurements. Collisions  between SDSS fibers or the  finite number of
science  fibers on  the SDSS  spectroscopic plates  mean that  not all
candidate  sources  for  follow-up  spectroscopy  can  be  assigned  a
fiber. Moreover, for the sources  that are observed the rate of secure
redshift measurement  depends on their optical  brightness. At fainter
magnitudes the signal-to-noise ratio  of the optical spectra decreases
and therefore the ability to estimate redshifts is affected.   The
probability of  a source being assigned  a fiber is  random, while the
redshift success rate depends, at least to the first approximation, on
optical  magnitude.     The   factor  $\eta(r)$  is  the  number  of
spectroscopically  confirmed X-ray sources  in the  magnitude interval
$r\pm\Delta r$ divided  by the total number of  X-ray sources that are
potential  targets for  follow-up spectroscopy  in the  same magnitude
range.  The  parent X-ray sample  of potential targets is  selected to
have  $f_X(\rm 0.5-10\,keV,  \Gamma=1.4)>10^{-14}\, erg  \,  s^{-1} \,
cm^{-2}$  and $15<r<22.5$\,mag (see  Section~\ref{sec_data_xxl}).  The
$r$-band   magnitude    dependence   of   $\eta(r)$    is   shown   in
Figure~\ref{fig_zeff_hist}.     For    magnitudes    in   the    range
$r=17.0-21.5$\,mag the efficiency  factor $\eta(r)$ is nearly constant
and larger than 80\%.  This  fraction drops to about 50\% at $r=22.5$,
the  limiting  magnitude for  follow-up  spectroscopy  with the  Sloan
telescope, and is zero for $r>22.5$\,mag.

 In  equation~\ref{eq_xxl} for  the  calculation  of the  optical
k-corrections  we use  the simulated  QSO SEDs  of \cite{McGreer2013}.
They  are  generated assuming  a  double  power-law  continuum in  the
rest-frame  UV/optical part  of the  SED  with a  break-point at  $\rm
1100\,  \AA$.  The short-  and long-wavelength  slopes are  drawn from
normal distributions  with means  of $-1.7$ and  $-0.5$, respectively.
For  both slopes the  scatter is  fixed to  0.3.  Emission  lines with
luminosity  dependent   equivalent  widths  as  well   as  Lya  forest
absorption  are also  added  to  the simulated  SEDs  as described  in
\cite{McGreer2013}. A total of 60\,000 model SEDs are generated in the
redshift interval  $z=3-5$ and for  AGN luminosities $\log  L_\nu (\rm
2500\,\AA)  \approx 29  - 32  \, erg  \, s^{-1}$.   These are  used to
calculate the  expected distribution of the  observed $r$-band optical
magnitudes for a given redshift  and AGN luminosity, and determine the
term $B(L_\mathrm{X},z,  \;|\; r)$ in  equation equation~\ref{eq_xxl}.
For the calculation of X-ray k-corrections we assume a power-law X-ray
spectrum with $\Gamma=1.9$.

\begin{figure*}
\begin{center}
\includegraphics[height=0.8\columnwidth]{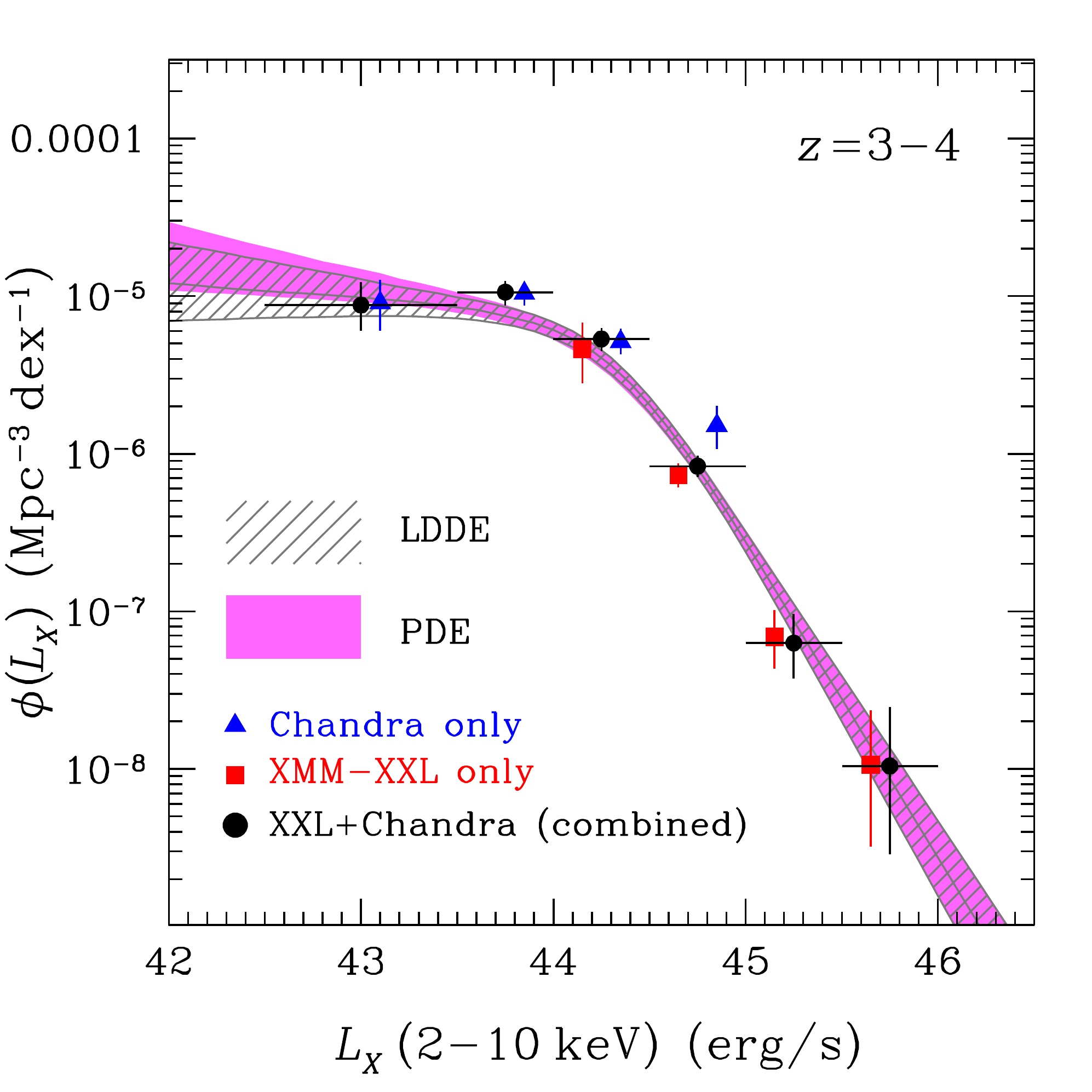}
\includegraphics[height=0.8\columnwidth]{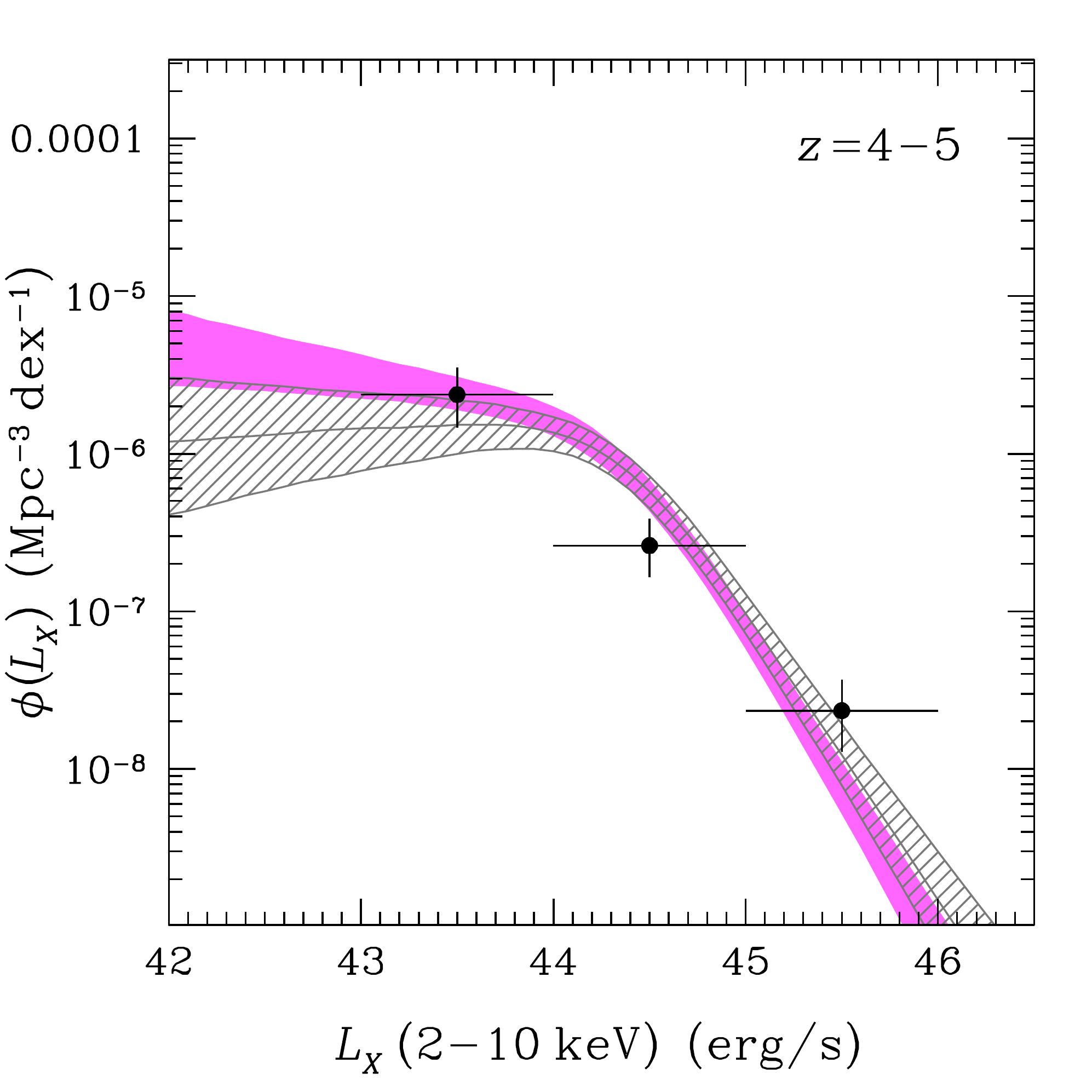}
\end{center}
\caption{AGN space density as a function of 2-10\,keV X-ray luminosity
in the redshift intervals $z=3-4$ and $z=4-5$.  In both panels the
datapoints are  the non-parametric  binned X-ray luminosity
function. The  
errors  correspond   to  the  68th  percentiles   of  the  Probability
Distribution Function. The black  filled circles are the estimates for
the combined XMM-XXL  and Chandra deep field surveys.   In the $z=3-4$
panel (left) we also plot the constraints obtained separately from the
XMM-XXL  (red  squares)  and  the  Chandra deep  survey  fields  (blue
triangles).   This is  to demonstrate  that in  overlapping luminosity
bins  the XMM-XXL  and  Chandra data  yield  consistent results.   The
shaded regions are  the 68\% confidence intervals of  the LDDE and PDE
parametric models described in Section~\ref{sec_xlf_models}.  The gray
hatched region is for the LDDE and the pink region is for the PDE. The
parametric models are estimated at the middle redshift of the redshift
intervals, i.e.  $z=3.5$ and $z=4.5$.  }\label{fig_xlf}
\end{figure*}

\subsection{X-ray luminosity function models}\label{sec_xlf_models}

The Bayesian framework outlined above explicitly requires a model with
a set of free parameters  that are constrained by the observations. We
consider  both  non-parametric and  parametric  models  for the  X-ray
luminosity function in the redshift range $z=3-5$.  The non-parametric
model simply assumes that the  space density of AGN is constant within
a given luminosity and redshift interval.  In this case the free model
parameters  to  be estimated  are  the  AGN  space densities  in  each
luminosity and  redshift bin.  This  is equivalent to the  widely used
$\rm   1/V_{max}$\footnote{In   the    case   of   a   step   function
(i.e.   non-parametric)  model  for   $\phi(L_X,  z)$   and  vanishing
uncertainties  for  the   redshifts  and  luminosities  of  individual
sources, it is straightforward to show that the maximum likelihood value of
$\phi(L_X, z)$ in equation  1 reduces to the  \protect\cite{Page_Carrera2000}
binned luminosity  function estimator.}approach for  the determination
of binned  luminosity functions.  The advantage  of the non-parametric
approach is that it allows  investigation of the form and amplitude of
the X-ray luminosity function evolution in a model-independent way.

We  also consider  four  parametric models  for  the X-ray  luminosity
function and its redshift  evolution, which have been extensively used
in the literature. These are the Pure Luminosity Evolution (PLE), Pure
Density Evolution (PDE), Luminosity Dependent Density Evolution (LDDE)
and   Luminosity   And    Density   Evolution   (LADE)   models.   The
parametrisation  of  each  model  follows \cite{Vito2014}.  The  X-ray
luminosity function  in the redshift  range $z=3-5$ is defined  as the
space density of  AGN per logarithmic luminosity bin  and is described
by a double power-law of the form

\begin{equation}\label{eq_xlf} 
\phi(L_X,         z)        =        \frac{         K        }{
  \Big(\frac{L_X}{L_\star}\Big)^{\gamma_1} + \Big(\frac{L_X}{L_\star}\Big)^{\gamma_2} },
\end{equation}

\noindent  where  $K$  is  the  normalization,  $\gamma_1$  and
$\gamma_2$   are   the   faint   and  bright-end   power-law   slopes,
respectively, and  $L_\star$ is the  break luminosity.  The  PLE model
assumes that $L_\star$ is a function of redshift and evolves according
to the relation

\begin{equation}\label{eq_ple}  L_\star  (z)  = L_\star  (z_0)  \times
\Big ( \frac{1+z}{1+z_{0}} \Big ) ^{p},
\end{equation}

\noindent where we  fix $z_0=3$ and the exponent  $p$ parametrises how
fast the break luminosity evolves with redshift. In the case of PDE it
is  assumed that  only the  normalisation of  the  luminosity function
evolves with the redshift according to the relation

\begin{equation}\label{eq_pde}  K(z)  = K(z_0) \times
\Big( \frac{1+z}{1+z_{0}} \Big ) ^{q},
\end{equation}

\noindent where $z_0=3$ and the exponent $q$ parametrises the speed of
the normalisation  factor evolution.  The  LADE model adopted  here is
similar  to the  Independent Luminosity  and Density  Evolution (ILDE)
model described by \cite{Yencho2009}.  We use equation~\ref{eq_ple} to
parametrise  the  redshift  evolution  of  $L_\star$ and  also  add  a
normalisation evolution term of the form

\begin{equation}\label{eq_lade} K(z) = K (z_0) \times
10^{q\,(z-z_0)}.
\end{equation}

\noindent   Finally   we   consider   the  LDDE   parametrisation   of
\cite{Hasinger2005},    where     the    normalisation    factor    of
equation~\ref{eq_xlf} changes with redshift as

\begin{equation}\label{eq_ldde} K(z) = K(z_0) \times
\Big( \frac{1+z}{1+z_{0}} \Big ) ^ {q+\beta\,(\log L_X - 44)},
\end{equation}

\noindent  where as  previously $z_0=3$  and the  rate of  the density
evolution  also depends  on  the X-ray  luminosity  via the  parameter
$\beta$.

Recall  that in  addition to  the model  parameters that  describe the
X-ray luminosity function in the redshift range $z=3-5$ the likelihood
function   in  equation~\ref{eq_likeli}   also  includes   terms  that
correspond to the  total AGN space density outside  the redshift range
of  interest,  $z<3$ or  $z>5$  (see equation~\ref{eq_split_xlf}).   A
non-parametric  model  is adopted  to  describe  that  term.  This  to
minimise the impact of parametric model assumptions to the results and
conclusions. We  use 3  redshift bins in  the range $z=0-3$  with size
$\Delta  z=1$.   Each  redshift   bin  is  split  into  5  logarithmic
luminosity bins  in the  range $\log  L_X(\rm 2-10\,keV) =  41 -  46 $
(units  of $\rm erg\,s^{-1}$)  with width  $\Delta \log  L_X =1$\,dex.
One additional  term is used to  model the luminosity  function in the
range $z=5-6$.   A constant  AGN space density  is assumed  within the
above luminosity and redshift intervals.   We therefore use a total of
16 nuisance parameters  to model the AGN X-ray  luminosity function at
$z<3$ or $z>5$.

The MultiNest  multimodal nested sampling  algorithm \citep{Feroz2008,
  Feroz2009} is  used for both  Bayesian parameter estimation  and the
calculation of  the Bayesian  evidence, $\mathcal{Z}$, of  each model,
i.e. the  integral of  the model likelihood  over the  parameter space
allowed  by  the priors.  The  Bayesian  evidence  is used  for  model
comparison, i.e.  to select among different models the one that better
describes the data.

\subsection{Potential obscuration effects}

The  analysis presented  in  this paper  uses  the full-band  selected
sample to  determine the  X-ray luminosity function  of AGN  at $z>3$.
This  is because  of the  higher sensitivity  of the  0.5-10\,keV band
compared to  e.g.  the  2-10\,keV band, which  translates to  a larger
number of  sources. A  potential issue however,  is that  the analysis
described above ignores the impact  of obscuration on the observed AGN
flux.   Figure~\ref{fig_suppression}  shows  that  for  AGN  at  $z>3$
moderate  line-of-sight  columns, $\log  N_H  =  23$ ($\rm  cm^{-2}$),
suppress the observed 0.5-10\,keV flux of AGN by less than about 25\%.
Higher  column  densities  however,   have  a  larger  impact  on  the
0.5-10\,keV flux and therefore  result in incompleteness in the sample
that  is not  accounted for  in  our analysis.   Nevertheless, we  are
primarily  interested in  the  differential evolution  of  AGN in  the
redshift intervals  $z=3-4$ and $z=4-5$.  Figure~\ref{fig_suppression}
shows that for fixed obscuration and intrinsic AGN luminosity there is
little difference  in the level of flux  suppression between redshifts
$z=3$ and  $z=5$. Incompleteness  related to obscuration  is therefore
expected to be similar across the redshift range $z=3-5$.  This allows
direct comparison of the inferred  AGN space densities in the redshift
bins $z=3-4$  and $z=4-5$, under the assumption  that the distribution
of  AGN in  obscuration  does not  change  dramatically between  these
redshift   intervals.      Recent  studies   on   the  obscuration
distribution of AGN  \citep{Ueda2014, Buchner2015, Aird2015} show that
the  obscured  AGN  fraction  increases  with redshift,  at  least  to
$z\approx3$.  In \cite{Buchner2015} for example, the obscured fraction
in Compton  thin AGN is about 50\%  at $z<1$ and increases  to 70\% at
$z\approx3$. At  higher redshifts however, current  constraints on the
evolution  of the  obscured AGN  fraction are  still limited  by small
number  statistics.   Nevertheless,  there  are suggestions  that  the
obscured  AGN  fraction  remains  roughly constant  with  redshift  at
$z\ga3$ \citep[][]{Buchner2015}.

Obscuration related  effects potentially have  a larger impact  on the
analysis  of the  XMM-XXL data.   For these  sources it  is explicitly
assumed that  their optical/UV continua are not  obscured or reddened.
This  assumption  allows  us  to   place  them  on  the  $L_\nu  {(\rm
2\,keV)}-L_\nu (\rm 2500\,\AA)$ correlation for type-I AGN and correct
for the spectroscopic magnitude cut  as explained above.  This poses a
problem when comparing the Chandra with the XMM-XXL constraints on the
X-ray luminosity  function.  Nevertheless, the  shallow XMM-XXL survey
is only  sensitive to powerful  QSOs [$L_X(  \rm 2 -  10 \, keV  ) \ga
10^{44} \, erg  \, s^{-1}$] at $z>3$.  At lower  redshift at least, it
is  well  established  that  the  obscured  AGN  fraction  is  rapidly
decreasing       with       increasing      accretion       luminosity
\citep[e.g.][]{Ueda2003,     Akylas2006,     Ueda2014,    Buchner2014,
Merloni2014}.   Obscuration incompleteness  corrections  are therefore
likely to  be small for  luminosities $L_X(  \rm 2 -  10 \, keV  ) \ga
10^{44} \, erg \, s^{-1}$.   We explore this assumption further in the
next section by directly comparing the Chandra and XMM-XXL constraints
in overlapping luminosity bins.

\begin{figure}
\begin{center}
\includegraphics[height=0.8\columnwidth]{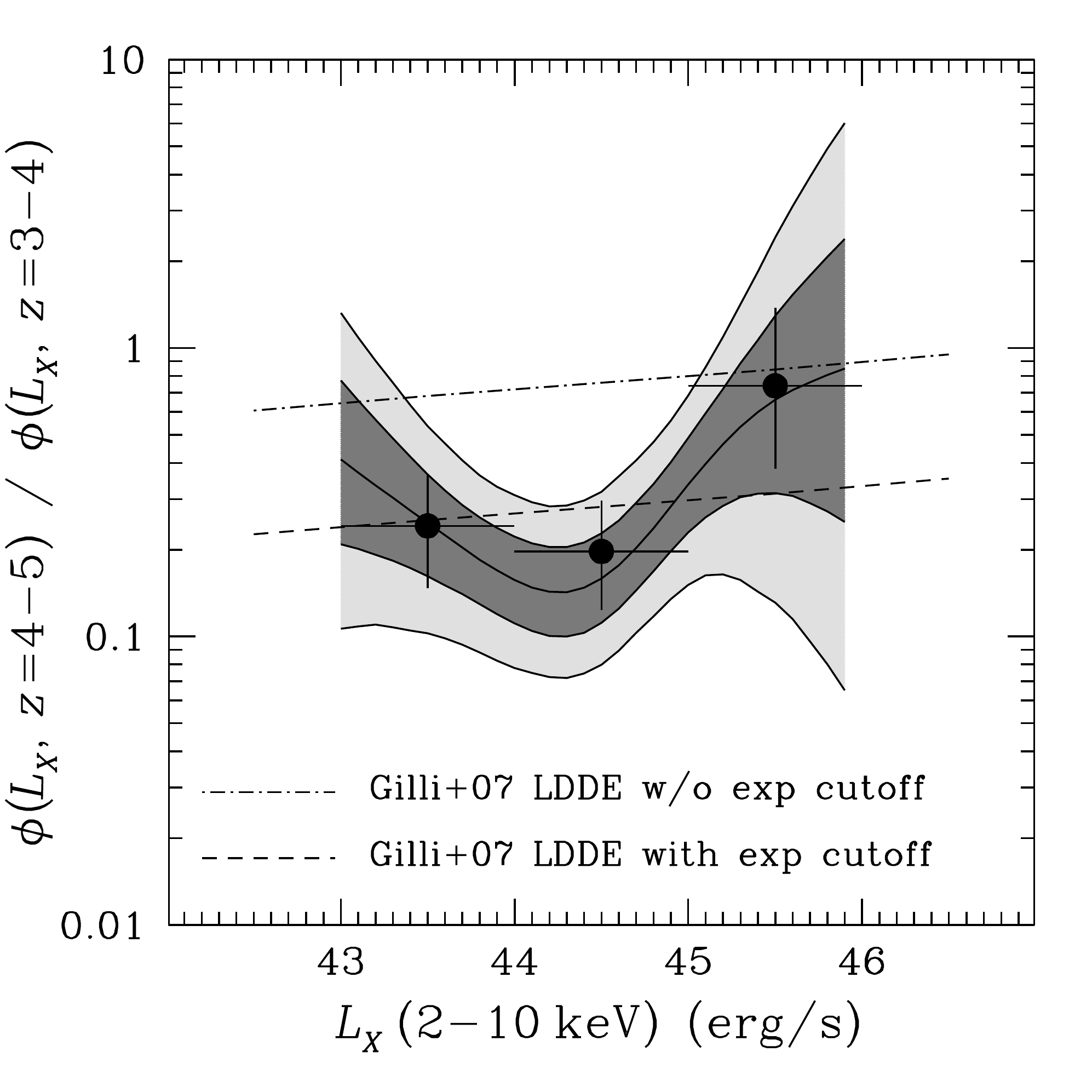}
\end{center}
\caption{X-ray  luminosity dependence of  the ratio  of the  AGN space
densities  in the redshift  intervals $z=3-4$  and $z=4-5$.   The data
points are for the  non-parametric binned X-ray luminosity function. A
double power-law is also fit independently to the data in the redshift
intervals $z=3-4$ and $z=4-5$.  The  ratio of these parametric fits is
shown  by the  light/dark  grey  shaded regions.   The  widths of  the
dark-grey  and  light-grey  shaded   regions  at  a  given  luminosity
corresponds to the 68\% and 95\% confidence intervals. The predictions
of  the Gilli  et al.  (2007) LDDE  parametrisation of  the luminosity
function  with   (dashed  line)  and  without   (dot-dashed  line)  an
exponential cutoff  at high redshift  is also plotted  for comparison.
}\label{fig_ratio}
\end{figure}

\begin{table}
\caption{Power-law fits to the data in the redshift intervals $z=3-4$
  and $z=4-5$}\label{tab_pl_fit}

\begin{tabular}{c cccc}

\hline

redshift & $\log K$  & $\log L_\star$ & $\gamma_1$ &  $\gamma_2$  \\
interval & ($\rm Mpc^{-3}$) & (erg/s)        &            &             \\

 (1) & (2) & (3) & (4) & (5) \\
\hline
$z=3-4$   & $-5.15_{-0.17}^{+0.12}$ & $44.41_{-0.12}^{+0.13}$ & $0.19_{-0.13}^{+0.15}$ & $2.25_{-0.24}^{+0.29}$ \\

$z=4-5$   & $-6.62_{-0.82}^{+0.74}$ & $45.00_{-0.58}^{+0.57}$ & $0.68_{-0.38}^{+0.24}$ & $2.16_{-0.47}^{+0.98}$ \\

\hline

\end{tabular}
\begin{list}{}{}
\item 
Listed are  the best-fit parameters  for the simple power-law  fits to
the data in the redshift intervals $z=3-4$, $z=4-5$. The listed values
are  the  median of  the  probability  distribution function  of  each
parameter.  The  errors correspond  to the  16th and  84th percentiles
around the  median. The columns  are: (1) redshift interval  (2) X-ray
luminosity  function  normalisation (see  equation~\ref{eq_xlf}),  (3)
break   luminosity    of   the   X-ray   luminosity    function   (see
equation~\ref{eq_xlf}), (4) faint-end slope, (5) bright-end slope.
\end{list}
\end{table}

\section{Results}

\subsection{The non-parametric X-ray luminosity function determination}

Figure~\ref{fig_xlf}  plots the non-parametric  estimate of  the X-ray
luminosity function  in two  redshift intervals, $z=3-4$  and $z=4-5$.
For the low redshift sub-sample,  $z=3-4$, we also plot separately the
AGN  space  density determined  independently  from  the Chandra  deep
fields  and the  XMM-XXL  survey.  As  expected  these datasets  probe
different luminosity  ranges.  The Chandra  fields provide constraints
to $\log  L_X(\rm 2-10\,keV)=42.5$  ($\rm erg \,  s^{-1}$, but  do not
probe  sufficient volume  to  detect luminous  and  rare sources  with
luminosities $\rm \ga  10^{45} \, erg \, s^{-1}$.   The XMM-XXL survey
can constrain  the bright-end  of the luminosity  function but  is not
sensitive enough  to detect  AGN below $L_X(\rm  2-10\,keV)\la 10^{44}
\rm \,  erg \,  s^{-1}$. Nevertheless, in  the interval  $\log L_X(\rm
2-10\,keV)=44-45$ ($\rm erg \, s^{-1}$)  both the Chandra and XMM data
provide meaningful  constraints to the AGN space  density.   These
independent non-parametric estimates are  in good agreement within the
68\%  errors  plotted in  Figure~\ref{fig_xlf}.   This suggests  that
obscuration  effects do  not have  a strong  impact on  the  AGN space
densities  estimated from the  XMM-XXL data.   In Figure~\ref{fig_xlf}
the $z=4-5$ panel does not plot separately the space density estimates
from the Chandra  and then XMM-XXL survey fields.   This is because at
this  redshift   interval  there  is  virtually  no   overlap  in  the
luminosities probed  by the two datasets. The  XMM-XXL constrains only
the brightest  luminosity bin, $\log  L_X(\rm 2-10\,keV)=45-46$ (units
$\rm erg  \, s^{-1}$).  The  Chandra fields include only  sources that
are fainter than $L_X(\rm 2-10\,keV)=10^{45} \, erg \, s^{-1}$ .

A striking  result illustrated  in Figure~\ref{fig_xlf} is  the strong
evolution  of   the  AGN   population  between  redshifts   $z=3$  and
$z=5$. This  is further demonstrated  in Figure~\ref{fig_ratio}, which
plots as a function of  luminosity the ratio of the non-parametric AGN
space  densities in the  redshift intervals  $z=3-4$ and  $z=4-5$.  At
luminosities $L_X(\rm 2-10\,keV)\la10^{45}\, erg \, s^{-1}$ there is a
factor of 5 decrease in the AGN space density from $z=3-4$ to $z=4-5$.
There is  also evidence for luminosity dependent  evolution.  AGN with
$L_X(\rm     2-10\,keV)\la10^{45}\,      erg     \,     s^{-1}$     in
Figure~\ref{fig_ratio}  appear  to  evolve faster  than  intrinsically
brighter sources. We quantify the statistical significance of the
evidence for luminosity dependent evolution using the quantity

\begin{equation}
\begin{split}
\mathcal{R} = & \log \frac{\phi(L_X=10^{45}-10^{46},
  z=4-5)}{\phi(L_X=10^{45}-10^{46}, z=3-4)} - \\
  & \log \frac{\phi(L_X=10^{43}-10^{45},
  z=4-5)}{\phi(L_X=10^{43}-10^{45}, z=3-4)},   
\end{split}
\end{equation} 

\noindent  where  $\phi(L_X, z)$  is the  binned (non-parametric)
luminosity  function.   $\mathcal{R}$  is the  logarithmic  difference
between the  high luminosity ($L_X>10^{45}\rm \, erg  \, s^{-1}$) data
point  of  Figure~\ref{fig_ratio} and  the  sum  of  the two  moderate
luminosity data  points ($L_X=\rm 10^{43} - 10^{45}\,  erg \, s^{-1}$)
of the same plot. We  bin together moderate luminosity AGN to simplify
the  problem  and also  improve  the  statistical  reliability of  the
results.   We  use  the   full  probability  density  distribution  of
$\mathcal{R}$ and  find that at  the 90\% probability  $\mathcal{R}>0$, i.e.
there is  differential evolution between moderate  ($L_X=\rm 10^{43} -
10^{45}\,  erg \,  s^{-1}$) and  powerful ($L_X>10^{45}\rm  \,  erg \,
s^{-1}$) X-ray  AGN from $z=3-4$ to $z=4-5$.   We further investigate
this using a simple  parametric approach.  A double power-law function
(equation~\ref{eq_xlf})  is  fit  independently  to the  data  in  the
redshift  bins $z=3-4$  and  $z=4-5$.  In  this exercise  evolutionary
effects within  each of the  two redshift intervals are  ignored.  The
best-fit parameters are presented in Table~\ref{tab_pl_fit}. The ratio
between the  two double  power-laws in the  redshift bins  $z=3-4$ and
$z=4-5$ is plotted with  the shaded region in Figure ~\ref{fig_ratio}.
The  apparent increase  of  this ratio  toward  faint luminosities  is
because  the   faint-end  slope  of  the  $z=4-5$   sample  is  poorly
constrained  and on  the  average  steeper than  that  of the  $z=3-4$
sample.  Nevertheless,  in the  interval $L_X  ( \rm 2-10  \, keV  ) =
10^{43}-10^{46} \, erg \, s^{-1}$,  where constraints on the AGN space
density  are   available  for  both   the  $z=3-4$  and   the  $z=4-5$
sub-samples, the  ratio between the two power-laws  is consistent with
luminosity-dependent  evolution. Larger  samples  are needed  however,
particularly  in   the  redshift  interval  $z=4-5$,   to  reduce  the
uncertainties  in  Figure~\ref{fig_ratio}   and  further  explore  the
evidence for luminosity dependent evolution.

The amplitude of  the AGN X-ray luminosity  function evolution between
the redshift  intervals $z=3-4$ and $z=4-5$  in Figure~\ref{fig_ratio}
is  further  compared  with  the predictions  of  luminosity  function
parametrisations  that  include  an  exponential  decline  at  $z\ga3$
\citep[e.g.][]{Gilli2007}.   Such models  are motivated  by the  rapid
evolution   of   optical   QSO   space  density   at   high   redshift
\citep[e.g.][]{Schmidt1995, Richards2006}.   We use  the Gilli  et al.
(2007)\footnote{\url{http://www.bo.astro.it/~gilli/counts.html}}  LDDE
parametrisation of the  X-ray luminosity function with  and without an
additional exponential cutoff  to predict the space density  of AGN at
$z=3.5$ and $z=4.5$. The ratio between these predictions is plotted in
Figure~\ref{fig_ratio}. The amplitude of the AGN evolution inferred in
this  paper for  luminosities $L_X(\rm  2-10\,keV)\la10^{45}\, erg  \,
s^{-1}$   is  consistent   with  the   Gilli  et   al.   (2007)   LDDE
parametrisation  that  includes  an   exponential  cutoff.   AGN  with
luminosities in the  range $L_X ( \rm 2-10 \,  keV ) = 10^{45}-10^{46}
\,  erg \,  s^{-1}$  lie in  between  the Gilli  et  al. (2007)  model
predictions with and without an exponential cutoff. This is suggestive
of milder evolution, albeit at the $\approx90\%$ confidence level.

\subsection{Parametric X-ray luminosity function determination}

Next  we  use  Bayesian  model  comparison  to  assess  which  of  the
evolutionary models  outlined in Section~\ref{sec_xlf_models} provides
a  better description  of the  observations in  the  redshift interval
$z=3-5$.   The   PLE,  PDE,  LADE   and  LDDE  parametric   models  of
Section~\ref{sec_xlf_models}  are  fit  to  the combined  Chandra  and
XMM-XXL   dataset.   Table~\ref{tab_xlf_fit}  presents   the  best-fit
parameters for each parametric model.  The Bayes factor (ratio between
evidences)  of each  model  relative  to the  model  with the  highest
evidence (PDE) is also shown in that table.

 The model with the highest evidence in Table~\ref{tab_xlf_fit} is the
PDE, with the second best being  the LDDE. The Bayes factor of the two
models  is  $\Delta  \log_{10}  Z  = 0.25$.   Based  on  the  Jeffreys
interpretation   of  the   Bayes   factor  \citep{Jeffreys1961}   this
difference  suggests  that  both  models  describe  equally  well  the
evolution in the  redshift interval $z=3-5$ of the  X-ray selected AGN
sample presented  in this paper.  The  bulk of the AGN  in the present
sample have X-ray  luminosities $L_X ( \rm 2-10 \, keV  ) < 10^{45} \,
erg  \,  s^{-1}$.   Figure~\ref{fig_ratio}  shows  that  such  sources
experience similar  reduction in  their space density  between $z=3-4$
and  $z=4-5$,  i.e.  consistent  with  pure  density evolution.   More
luminous AGN [$L_X (  \rm 2-10 \, keV ) > 10^{45}  \, erg \, s^{-1}$],
which appear to experience milder evolution in Figure~\ref{fig_ratio},
represent only a small fraction  of the present sample.  This combined
with the fact that the PDE is  a simpler model compared to the LDDE (5
vs 6 free parameters) results  in similar Bayesian evidences for these
two parametric models.

Table~\ref{tab_xlf_fit}  also  shows  that  the  LADE  parametrisation
adopted in this  work performs worse than the  LDDE.  The Bayes factor
of the two models is $\Delta  \log_{10} Z = 1.08$.  This difference is
strong evidence in favor of the LDDE \cite{Jeffreys1961}.

The one model  that performs significantly worse than  the rest is the
PLE.  The  Bayes factor  of that  model relative to  the one  with the
highest evidence (PDE)  is $\Delta \log_{10} Z =  4.02$.  Based on the
Jeffreys interpretation of  the Bayes factor \citep{Jeffreys1961} this
is decisive evidence against the PLE model.

  Although the PDE model is favoured by our analysis for the evolution
of AGN  in the redshift interval  $z=3-5$ in the next  sections we use
the LDDE model  to compare with previous studies  and to determine the
contribution of X-ray  AGN to the ionisation of  the Universe. This is
because of  the small difference  in the evidences  of the PDE  vs the
LDDE  and  the  long   literature  on  the  LDDE  evolutionary  model.
Nevertheless,  the results and  conclusions are  not sensitive  to the
particular choice of AGN evolutionary model.

\subsection{Comparison with previous studies}

A  number of  studies on  the X-ray  luminosity function  of  AGN have
appeared in the literature  recently.  These include works focusing on
the   space  density   of  X-ray   AGN  at   high   redshifts  $z\ga3$
\citep{Civano2010,  Kalfountzou2014,  Vito2014}  and  results  on  the
global   X-ray   luminosity   function   evolution   across   redshift
\citep{Ueda2014, Buchner2015, Aird2015,  Miyaji2015}.  In this section
we compare  our X-ray luminosity function with  previous studies using
X-ray  samples  selected  in  the  0.5-2 or  0.5-10\,keV  bands,  i.e.
similar to  the one presented in this  paper. \cite{Vito2014} compiled
one of the  largest samples to date in  the redshift interval $z=3-5$.
They  select   sources  in  the  0.5-2\,keV  band   and  also  present
non-parametric  ($\rm 1/V_{max}$)  estimates of  the  X-ray luminosity
function. These  aspects of the  Vito et al. analysis  methodology are
similar to ours.   We also use recent results  of \cite{Aird2015} as a
representation of  parametric approaches  to fit the  X-ray luminosity
function of  AGN to the  full redshift interval accessible  to current
X-ray selected samples, $z\approx0-5$.

\begin{figure*}
\begin{center}
\includegraphics[height=0.8\columnwidth]{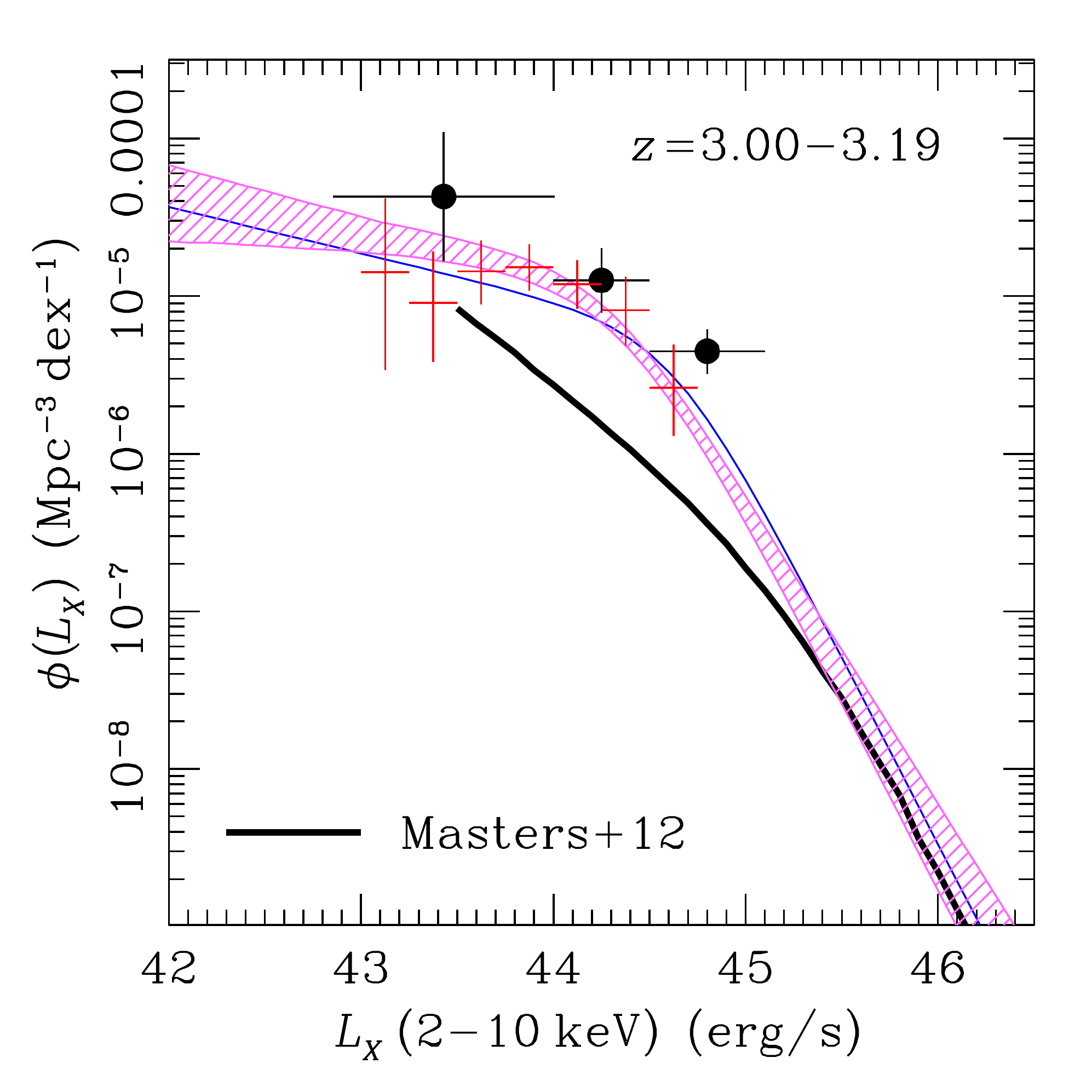}
\includegraphics[height=0.8\columnwidth]{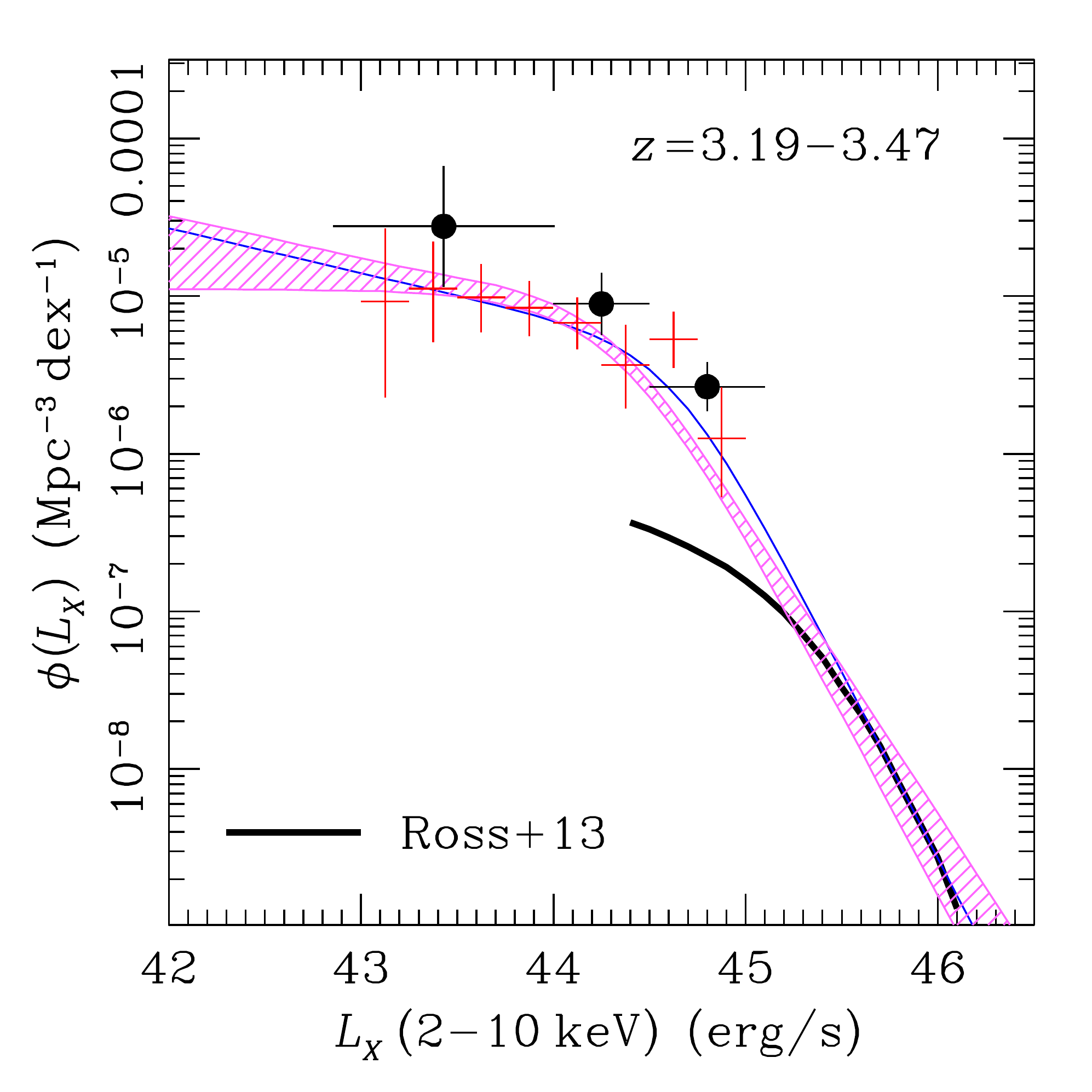}
\includegraphics[height=0.8\columnwidth]{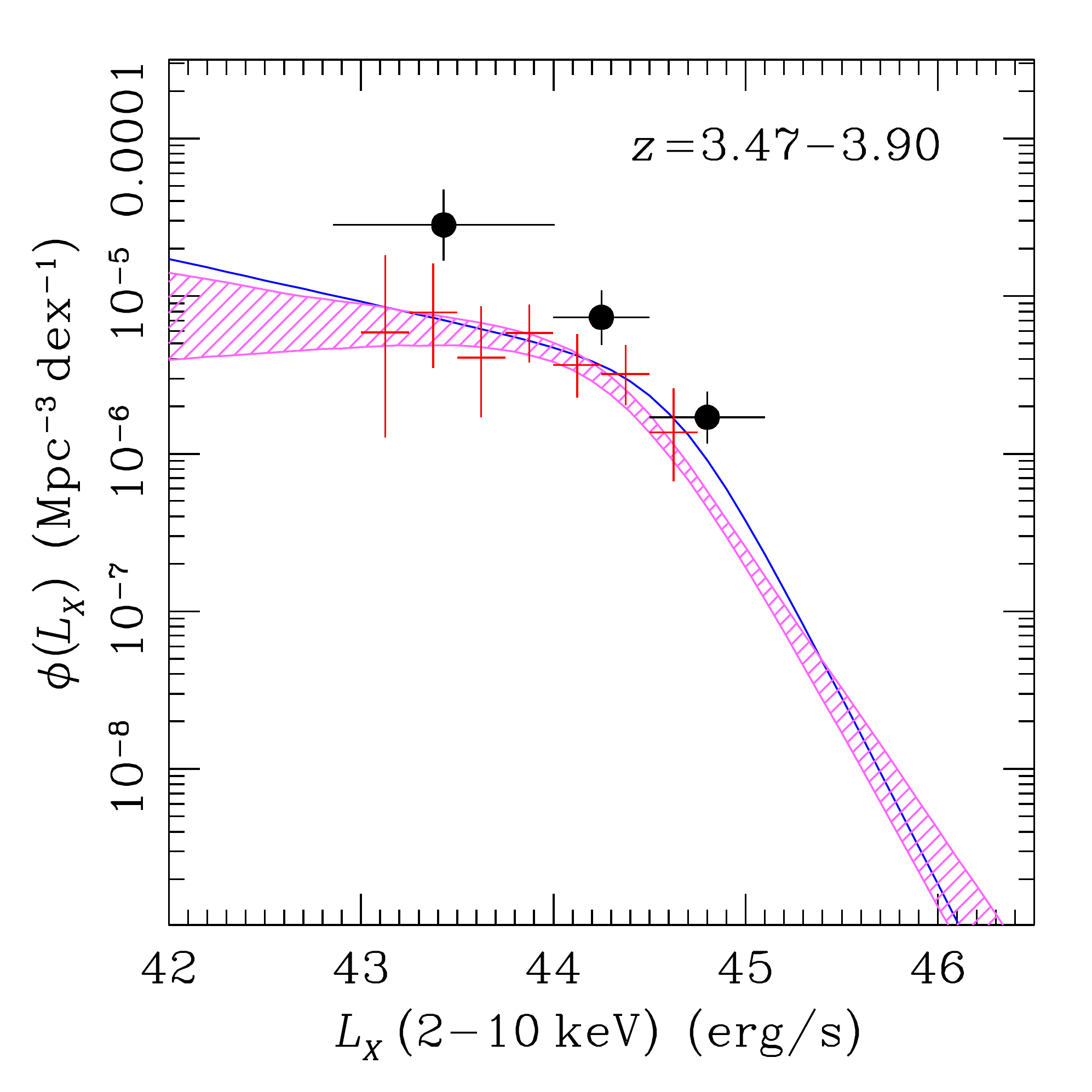}
\includegraphics[height=0.8\columnwidth]{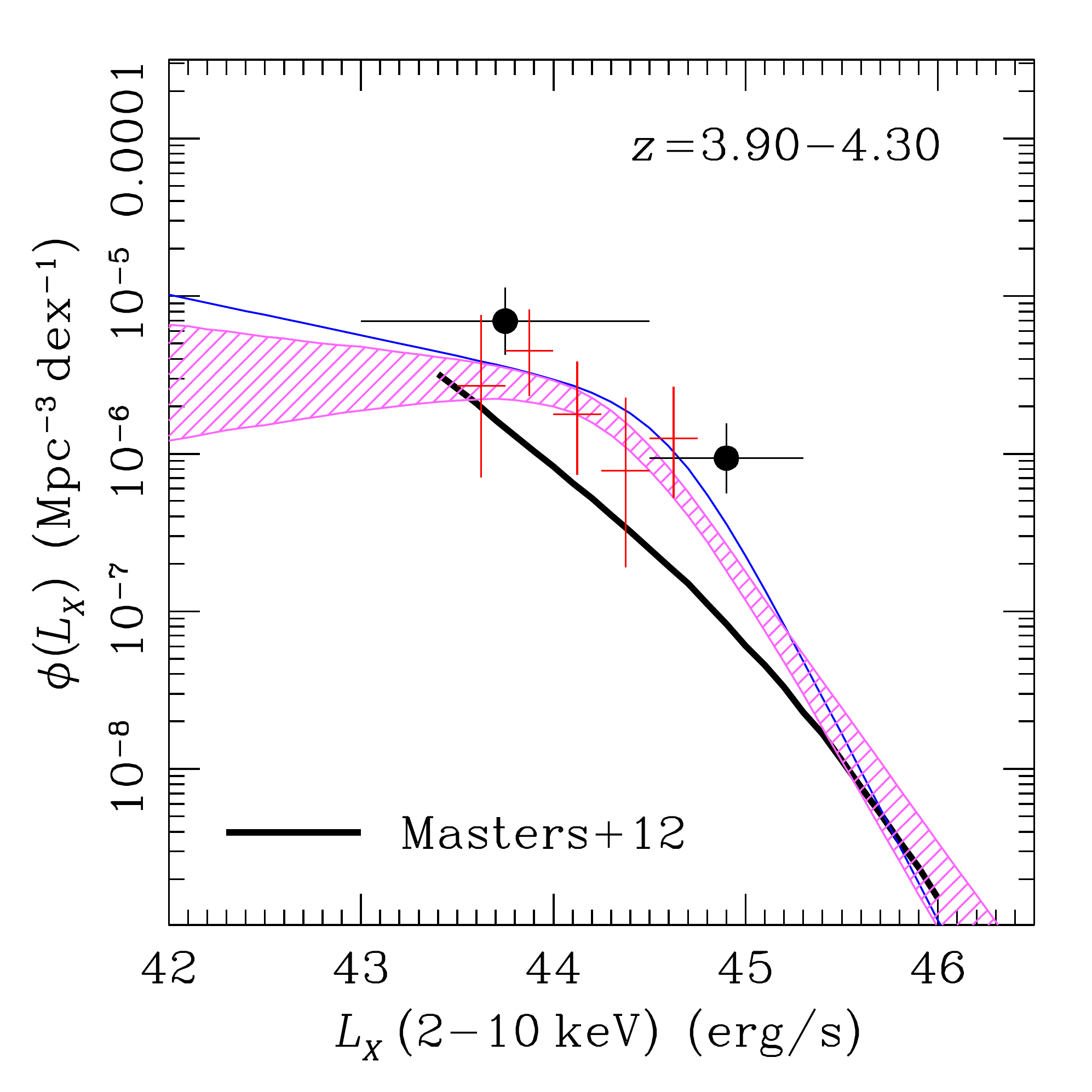}
\includegraphics[height=0.8\columnwidth]{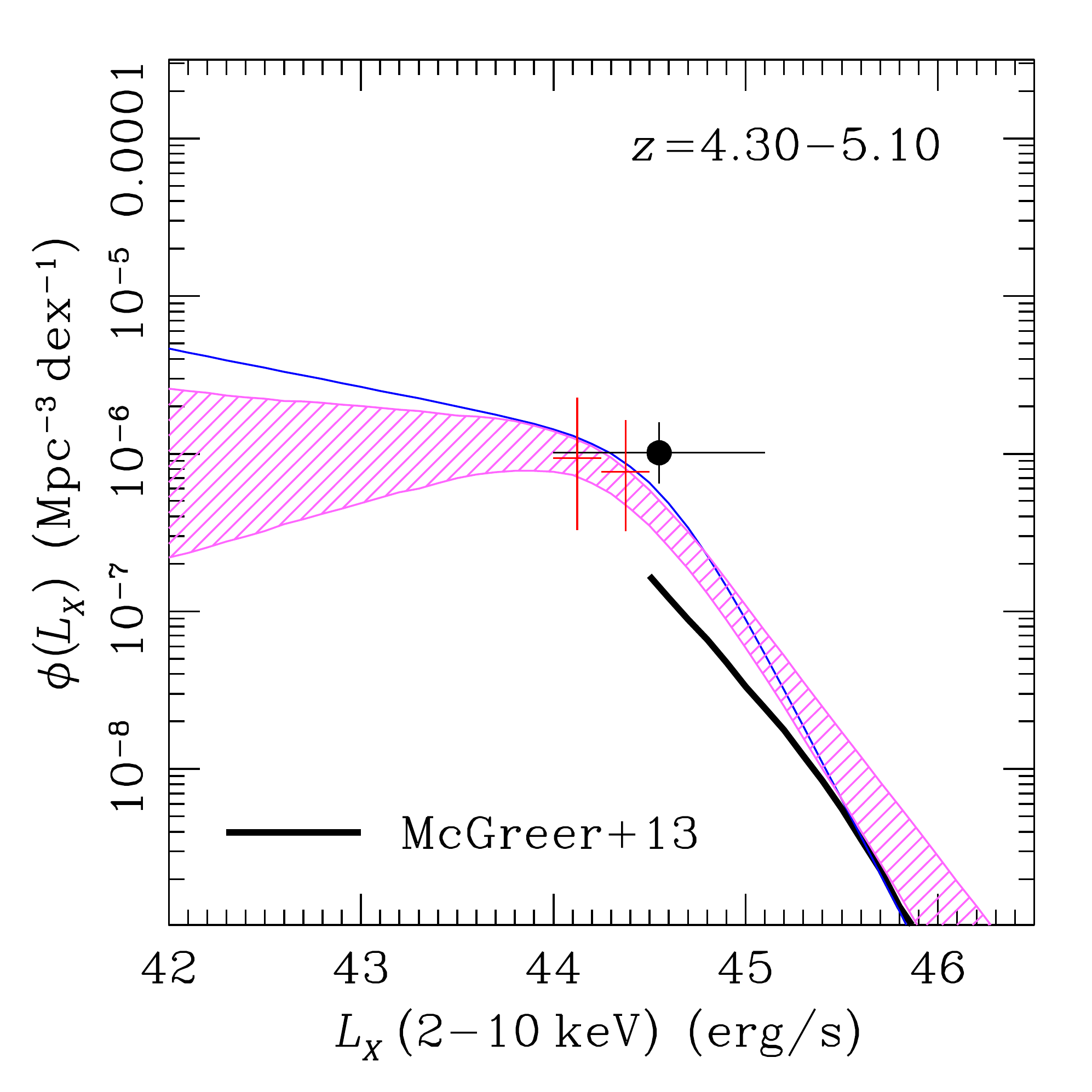}
\end{center}
\caption{Comparison of  our best-fit  LDDE model (pink  shaded region)
with previous estimates of the  X-ray luminosity function in the range
$z=3-5$.  The panels  correspond to the redshift intervals  of Vito et
al.  (2014) to  allow direct comparison with their  results. The black
filled  circles are  the  $\rm 1/V_{max}$  binned luminosity  function
estimates of  Vito et  al. (2014).  The  thin blue curves are  the flexible
double  power-law parametric  model of  Aird et  al. (2015)  for their
0.5-2\,keV band selected  sample without corrections for obscuration.
For  the  Aird  et  al.  (2015) 0.5-2\,keV  sample  the  binned  X-ray
luminosity function estimates are also  shown by the red crosses.  The
UV/optical QSO luminosity functions  of Masters et al. (2012), McGreer
et al.  (2013) and Ross  et al.  (2013)  are also plotted by  the thick
black lines at the relevant redshift intervals.  }\label{fig_xlf_vito}
\end{figure*}

Vito et al.  (2014) use a  sample similar to ours in size to determine
the X-ray luminosity  function in the range $3<z\la5$.   They use both
photometric  and  spectroscopic redshifts  (total  of  141) and  apply
corrections  to  account  for  sources  without  photometric  redshift
determinations (total of 65).  We compare our best-fit LDDE model with
their results in Figure~\ref{fig_xlf_vito}.  The redshift intervals of
each panel  are the same as  in Figure 7 of  \cite{Vito2014}.  We plot
their  non-parametric  $\rm 1/V_{max}$  estimates  that include  their
redshift  incompleteness corrections.  Our  LDDE parametric  models in
Figure~\ref{fig_xlf_vito} is estimated at  the middle of each redshift
interval.  We find that our X-ray luminosity function determination is
systematically lower than  the Vito et al.  (2014)  data points.  This
discrepancy  ($>3\sigma$ significance in  e.g.  the  redshift interval
$z=3.47-3.90$ of Figure~\ref{fig_xlf_vito}) may be related to the fact
that  Vito  et  al.   (2014)  account for  X-ray  obscuration  in  the
determination of  X-ray luminosities and  the calculation of  the $\rm
V_{max}$ of  individual sources.  A total of  36 of their  141 sources
have column densities  $\rm N_H\ga10^{23}\, cm^{-2}$, which translates
to  suppression  of their  fluxes  compared  to  the unobscured  ($\rm
N_H=0\,cm^{-2}$)     case     by     $\ga    20$\%     (see     Figure
\ref{fig_suppression}).  Such corrections  are ignored in the analysis
presented here.   Alternatively the discrepancy may be  related to how
sources  with photometric  redshifts or  sources without  any redshift
information  are  treated.   \cite{Vito2014}  use  only  the  best-fit
photometric  redshift   solution  without  taking   into  account  the
corresponding  uncertainties.   It  is  further assumed  that  the  65
sources in their  sample without photometric redshifts all  lie in the
redshift  interval   $z=3-5$  and   that  they  follow   the  redshift
distribution  of  the   X-ray  sources  with  redshift  determinations
(photometric or spectroscopic) in the range $z=3-5$.  The amplitude of
this  correction   is  $\approx+0.5$  and   $+0.25$\,dex  increase  at
luminosities  $L_X (\rm  2  - 10  \,  keV) \approx  10^{43}$ and  $\rm
10^{45}\, erg \,s^{-1}$ respectively.  In that respect \cite{Vito2014}
determine maximal X-ray luminosity  functions.  It is likely that some
of  the X-ray  sources without  redshift information  lie  outside the
redshift range $z=3-5$.   Heavily obscured AGN or AGN  hosted by early
type  hosts at moderate  redshifts, $z\approx1-3$,  may also  have red
SEDs, similar to those  of high redshift sources \citep{Koekemoer2004,
Schaerer2007, Rodighiero2007, DelMoro2009}.

\begin{figure*}
\begin{center}
\includegraphics[height=0.8\columnwidth]{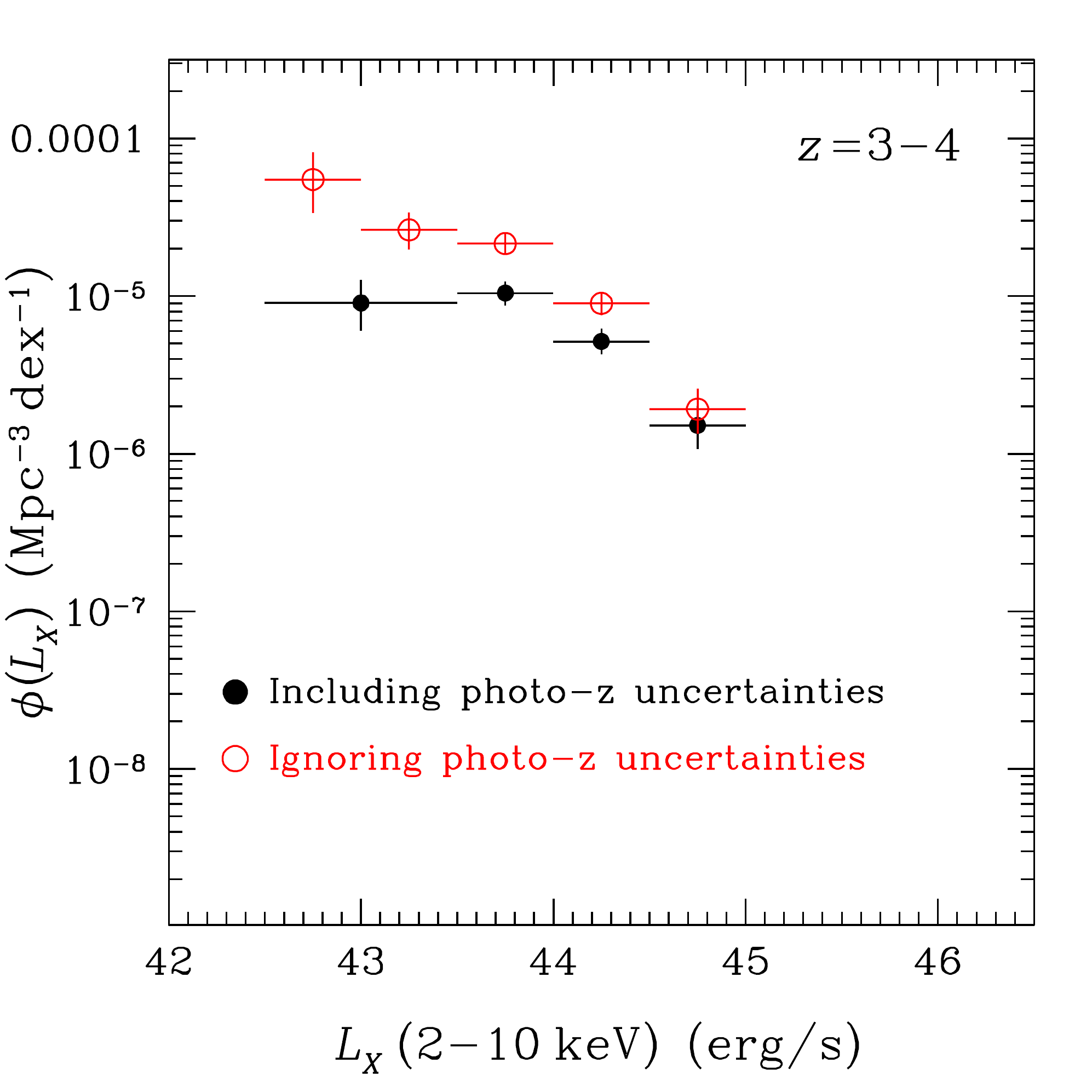}
\includegraphics[height=0.8\columnwidth]{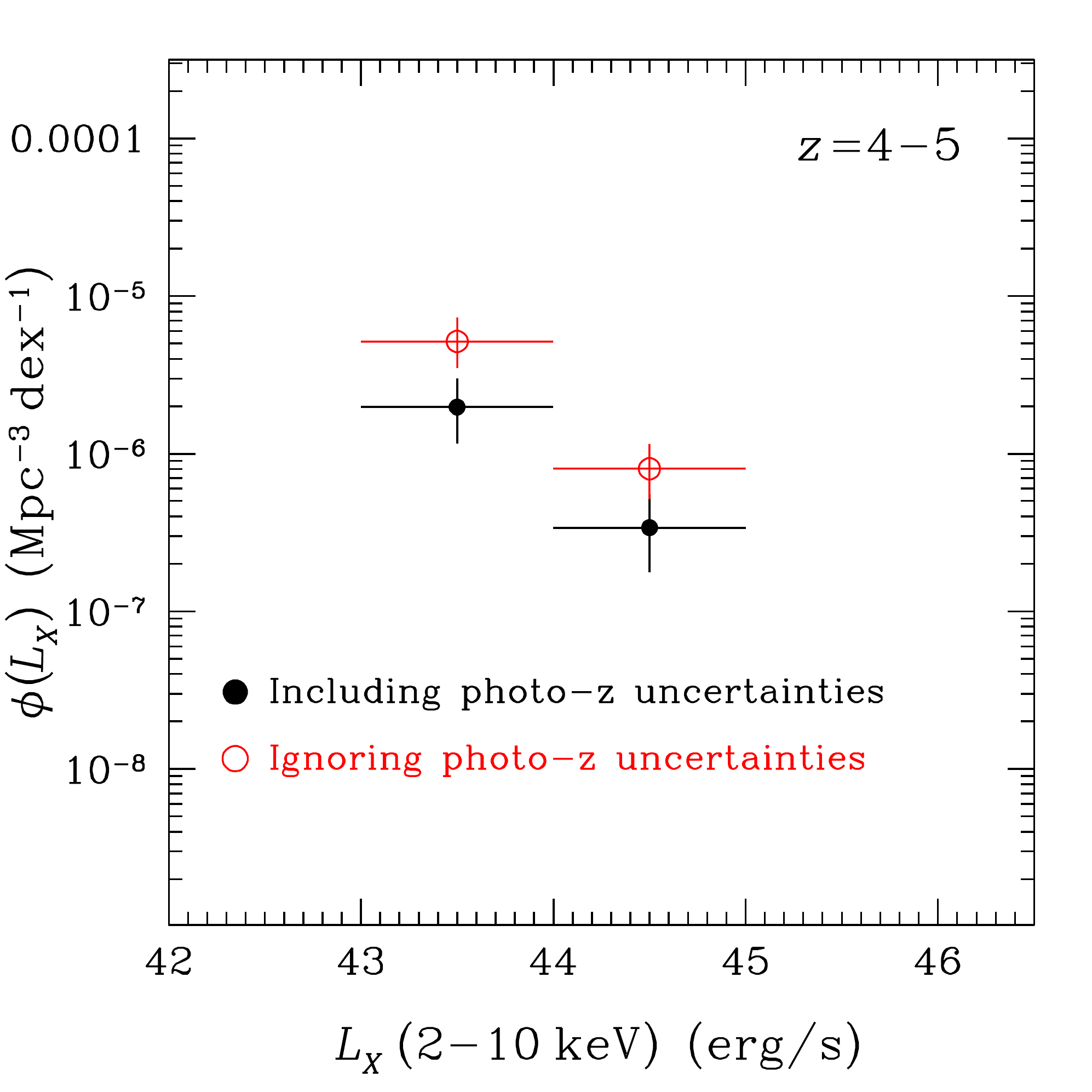}
\end{center}
\caption{Impact of ignoring photometric  redshift uncertainties on the
  inferred AGN space  density. The red open circles  correspond to the
  non-parametric binned  X-ray luminosity function estimated  by using
  the  best-fit  photometric  redshift   solution  only  and  ignoring
  photometric redshift uncertainties (see text for details). The black
  data points correspond  to the AGN space density  estimated by using
  the full  photometric redshift Probability Distribution  Function to
  account for photometric redshift errors. For both set of data-points
  only the Chandra survey fields are used.  }\label{fig_nophotoz}
\end{figure*}

We further investigate this issue by adopting a methodology similar to
that of  \cite{Vito2014} to  determine the X-ray  luminosity function.
We  use the best-fit  photometric redshifts  only, i.e.   ignoring the
photometric  redshift  probability  distribution  functions.   Sources
without  optical identifications  in  the sample  are assigned  random
redshifts in the  range $z=3-5$ based on the  redshift distribution of
sources with  spectroscopic or photometric  redshift measurements. The
results are presented in  Figure~\ref{fig_nophotoz}.  It shows that an
approach  that  ignores  photometric  redshift errors  results  in  an
overestimation of the AGN space  density.  This is not only because of
sources   without  optical   identifications  and   therefore  without
photometric  redshift (total  of 107  in our  sample)  estimates being
forced to lie in the  range $z=3-5$.  The overestimation is mainly the
result  of  ignoring   photometric  redshift  uncertainties.   In  the
likelihood equation~\ref{eq_likeli} the probability of a source having
redshift  $z$  and luminosity  $L_X$  is  weighted  by the  luminosity
function.   Sources  with   broad  (see  Fig.~\ref{fig_pdz})  redshift
probability  distribution functions  are  more likely  to  lie at  low
redshift and moderate luminosities simply because the space density of
AGN    is    higher     there.     Ignoring    this    weighting    in
equation~\ref{eq_likeli}  by e.g.  fixing  photometric redshifts  to a
single (best-fit) value overestimates  the luminosity function at high
redshifts.

\begin{figure}
\begin{center}
\includegraphics[height=0.8\columnwidth]{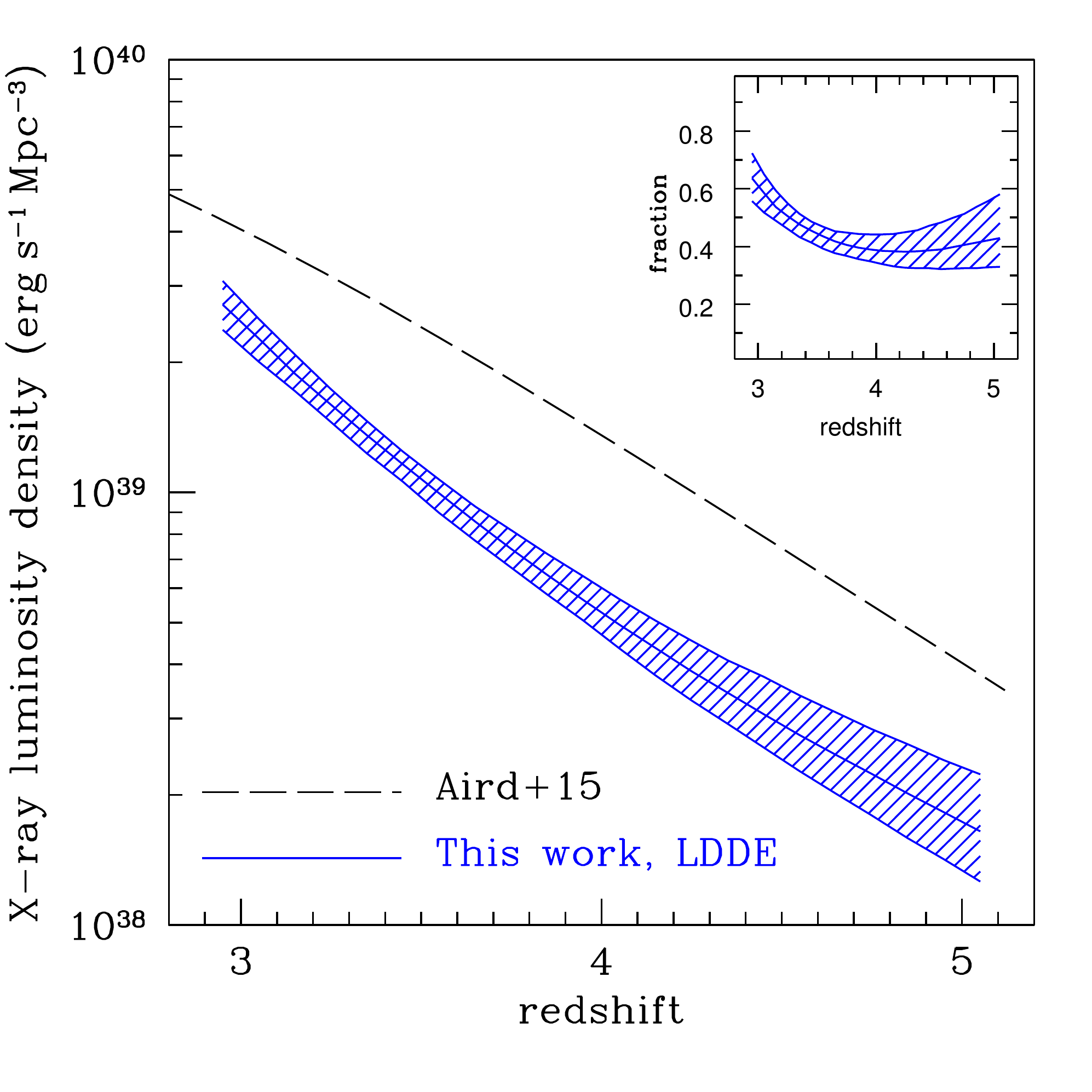}
\end{center}
\caption{X-ray luminosity density as a function of redshift. The black
  dashed curve corresponds to the total X-ray luminosity density for
  both obscured (Compton thin and Compton thick) and unobscured AGN
  deterined by Aird et al. (2015). The shaded blue region is the X-ray
  luminosity density estimated using our LDDE parametrisation for the
  evolution of the AGN population at $z>3$. The inset plot shows the ratio
  between the blue (solid) and black (dashed) curves as a function of
  redshift. It represents  the fraction of the luminosity density
  accounted by our LDDE model relative to the total luminosity density
  (corrected for obscuration effects)
  estimated by the  Aird et al. (2015).
}\label{fig_ld}
\end{figure}

Also  plotted in  Figure~\ref{fig_xlf_vito}  are  the flexible  double
power-law parametric model of \cite{Aird2015} for their 0.5-2 selected
sample  before applying  corrections  for  obscuration (i.e.  directly
comparable  to our  analysis).  The binned  X-ray luminosity  function
estimates of  \cite{Aird2015} are also  shown.  These data  points are
estimated   using   the    $N_{obs}/N_{mdl}$   method   developed   by
\cite{Miyaji2001} and are therefore  not independent of the underlying
parametric model plotted  in Figure~\ref{fig_xlf_vito}.  Nevertheless,
these   binned  estimates   provide  a   measure  of   the  associated
uncertainties  and can  be used  to identify  redshift and  luminosity
intervals  where the  parametric model  provides a  poorer fit  to the
data.  The \cite{Aird2015}  X-ray luminosity  function estimated  from
their  0.5-2\,keV  selected  sample  is in  fair  agreement  with  our
results. This may be not surprising  given the overlap between the two
datasets and the similar approaches.  
Figure \ref{fig_ld}  plots the  X-ray luminosity density  evolution of
AGN  using  our  LDDE   model  and  the  \cite{Aird2015}  total  X-ray
luminosity  function that  includes  both obscured  (Compton thin  and
thick) and unobscured sources. The X-ray luminosity function estimated
in this  work accounts  for about 40\%  of the total  X-ray luminosity
density determined by \cite{Aird2015}.

%The  Aird  et  al.   (2015)
%luminosity  function  based  on  their  2-7\,keV  selected  sample  is
%systematically higher than our estimates.  This indicates the presence
%of  obscured  AGN,  which  are  not accounted  for  in  our  analysis.
%Nevertheless,   at  bright   accretion  luminosities,   $\log  L_X(\rm
%2-10\,keV)  \ga  44.5\,  erg\,  s^{-1}$,   the  Aird  et  al.   (2015)
%0.5-2\,keV  and 2-7\,keV  selected samples  yield consistent  results.
%This  is  evidence against  a  large  population of  heavily  obscured
%sources at these luminosities and  supports our approach for analysing
%the XMM-XXL data  to constrain the bright-end of  the X-ray luminosity
%function.

Finally,  we  also  compare  in  Figure~\ref{fig_xlf_vito}  the  X-ray
luminosity function  with high redshift determinations  of the optical
QSO  luminosity  function.  The  conversion  from UV/optical  to
X-rays  ultimately depends  on the  scatter in  the  relations between
bolometric and  X-ray or UV luminosities.  There  are suggestions that
the bolometric--to--X-ray  luminosity ratio has a  larger scatter than
the  bolometric--to--UV luminosity ratio  \citep{Hopkins2007_bol}.  We
therefore convert from UV/optical to X-rays by convolving the best-fit
parametric  model of  UV/optical QSOs  from the  relevant publications
with the  $L_\nu ({\rm 2\,keV})  - L_\nu (\rm 2500\,\AA)$  relation of
\cite{Lusso2010}    assuming    a    scatter   of    0.4\,dex    (e.g.
Figure~\ref{fig_hist_l2500}).    This    calculation   also   requires
assumptions  on  the  shape of  the  UV  SED  of QSOs  at  wavelengths
$\lambda\ga1450$\,\AA.   We assume  a  power-law of  the form  $L(\nu)
\propto  \nu^{-0.5}$  \citep{vandenBerk2001,  Telfer2002}.  A  steeper
slope, $L(\nu)  \propto \nu^{-0.61}$ \citep{Lusso2015},  translates to
an  X-ray  luminosity  function  that  is offset  by  $\delta\log  \rm
L_{X}\approx-0.03$\,dex compared  to $L(\nu) \propto  \nu^{-0.5}$. At
the  $z=3.1$  and   $z=4.1$  panels  of  Figure~\ref{fig_xlf_vito}  we
overplot  the \cite{Masters2012} best-fit  double power-law  models at
$z=3.2$ and $z=4$. At  the $z=3.35$ panel of Figure~\ref{fig_xlf_vito}
we  show the  \cite{Ross2013} LEDE  (Luminosity Evolution  and Density
Evolution)    model   fit    to   the    BOSS   Stripe82    QSO   data
\citep{Palanque-Delabrouille2011}.  We choose  not to extrapolate this
model  past  redshift  $z=3.5$,  the  limiting redshift  of  the  BOSS
Stripe82  QSO  sample.  At  the  redshift  interval $4.3<z<5.1$  (mean
redshift $z=4.7$)  we transform to  X-rays the optical  QSO luminosity
function determined by \cite{McGreer2013}.  The optical
luminosity    functions   are   plotted    by   shaded    regions   in
Figure~\ref{fig_xlf_vito}.   The   shape  and  normalisation   of  the
UV/optical  QSO luminosity  functions in  Figure~\ref{fig_xlf_vito} at
luminosities $\rm \ga  10^{45} \, erg \, s^{-1}$  are sensitive to the
scatter of the $L_\nu ({\rm 2\,keV}) - L_\nu (\rm 2500\,\AA)$ relation
used in the convolution.

A  striking   result  from  Figure~\ref{fig_xlf_vito}   is  the  steep
faint-end  slope of the  \cite{Masters2012} best-fit  double power-law
models at  $z=3.2$ and  $z=4$ compared to  the faint-end slope  of the
X-ray luminosity  function.  The ratio between the  UV/optical and the
X-ray luminosity  functions is  a proxy of  the type-I  fraction among
AGN,  $\mathcal{F}_{\rm  type-I}$, and  can  be  used  to explore  the
luminosity dependence of this quantity  at high redshift.  In this
section we define type-I AGN as sources with blue UV/optical continua,
typical  of broad-line  QSOs.   We  use the  Masters  et al.   (2012)
luminosity  function  and  the  LDDE  parametrisation  for  the  X-ray
luminosity function  to determine $\mathcal{F}_{\rm  type-I}$ and plot
the results in the case of $z=3.2$ in Figure~\ref{fig_typeI_fraction}.
The luminosity  dependence of the $\mathcal{F}_{\rm  type-I}$ at $z=4$
is very similar and is not shown. We find evidence that the type-I AGN
fraction at $z>3$ is a  non-monotonic function of luminosity. There is
a  minimum  at $L_X  (\rm  2  - 10  \,  keV)  \approx  10^{44} \,  erg
\,s^{-1}$, followed by a steep increase toward fainter luminosities.

The      behaviour      of      $\mathcal{F}_{\rm     type-I}$      in
Figure~\ref{fig_typeI_fraction}  is opposite  to studies  that  find a
drop in the type-I or  X-ray unobscured ($\rm N_H<10^{22} \, cm^{-2}$)
AGN fraction  with decreasing luminosity or  equivalently that type-II
or  obscured ($\rm  N_H>10^{22}  \, cm^{-2}$)  AGN  dominate at  faint
luminosities   \citep{Ueda2003,  Akylas2006,   Ueda2014,  Merloni2014,
Aird2015}. Studies on the obscuration distribution of AGN in the local
Universe support  a picture where the obscured  AGN fraction increases
with decreasing luminosity. Nevertheless,  they also find evidence for
a turnover  (drop) of  the obscured AGN  fraction at very  faint X-ray
luminosities,   below   about  $\rm   10^{42}   \,   erg  \,   s^{-1}$
\citep[e.g][]{Burlon2011,       Brightman_Nandra2011}.       Recently,
\cite{Buchner2015} extended these results to higher redshift and found
evidence that the obscured  AGN fraction peaks at a redshift-dependent
luminosity and  then drops at both brighter  and fainter luminosities.
Figure~\ref{fig_typeI_fraction}  overplots the  obscured  AGN fraction
derived by Buchner et al.   (2015) in the redshift interval $z=2.7-4$.
For  this comparison  we assume  that $\mathcal{F}_{\rm  type-I}=  1 -
\mathcal{F}_{\rm obscured}$.  This  is a simplistic assumption because
the definition  of Type-I QSOs in  the case of  UV/optical samples and
unobscured/obscured  AGN  in X-ray  samples  like  in  Buchner et  al.
(2015) is different.  Nevertheless, to the first order the quantity $1
- \mathcal{F}_{\rm  obscured}$ should be  at least loosely  related to
$\mathcal{F}_{\rm type-I}$.   In Figure~\ref{fig_typeI_fraction} there
is  qualitative agreement  between the  Buchner  et al.   (2015) $1  -
\mathcal{F}_{\rm   obscured}$   parameter   and  our   definition   of
$\mathcal{F}_{\rm type-I}$.

 At  bright luminosities  the quantity  $\mathcal{F}_{\rm  type-I}$ in
Figure~\ref{fig_typeI_fraction} increases with increasing $L_X$.  This
in    agreement    with    previous    studies    based    on    X-ray
\citep[e.g.]{Ueda2014},    optical    \citep[e.g.][]{Simpson2005}   or
infrared    \citep[e.g.][]{Assef2013}   data.    At    the   brightest
luminosities probed by  our data, $L_X = \rm 10^{45}  - 10^{46} \, erg
\, s^{-1}$, the type-I fraction is about $\rm 75\pm25\%$.  This number
is relevant to the population of powerful QSOs (bolometric $L_{bol}\rm
\approx 10^{47}  \, erg \, s^{-1}$) with  reddened UV/optical continua
[extinction  $\rm E(B-V)  \approx 5$]  identified in  recent wide-area
infrared  surveys \citep{Stern2014}.  These  reddened/obscured sources
correspond to  X-ray luminosities $L_X(\rm  2-10\,keV) \approx 5\times
10^{45}  \, erg \,  s^{-1}$ \citep{Marconi2004}  and are  suggested to
represent  up to  50\% of  the  QSO population  at these  luminosities
\citep{Assef2014}.  We  caution however, that  there are uncertainties
on the  inferred bolometric luminosities  of these sources  and hence,
the obscured AGN fraction, depending on the assumed geometry, physical
scale    and   covering   fraction    of   the    obscuring   material
\citep{Assef2014}.   Additionally,  the   X-ray  properties  of  these
infrared selected  AGN are poorly  known.  There are  suggestions that
they represent heavily obscured,  even possibly Compton thick, systems
\citep{Stern2014},   in   which  case   they   are   expected  to   be
underrepresented in our sample.

\subsection{Contribution of AGN to ionisation of the Universe}

The  X-ray  luminosity function  can  be used  to  set  limits on  the
contribution of the AGN to the radiation field that keeps the Universe
ionised at redshift $z>3$. The advantage of X-ray selection is that it
provides  a better  handle  on  the faint-end  of  the AGN  luminosity
function compared  to UV/optically  selected samples. The  downside is
that assumptions on the escape fraction  of UV photons have to be made
to convert AGN  space densities to ionising photon  densities.  At the
very least however, X-rays surveys  can set strict upper limits on the
contribution of AGN to the ionising photon field, under the assumption
that  all  photons  emitted  by  the  central  source  escape  to  the
Inter-Galactic Medium.

The rate of hydrogen ionising  photons is estimated by integrating the
AGN Spectral Energy Distribution in the energy range 1-4\,ryd

\begin{equation} 
\dot{ n } = \int_{1ryd}^{4ryd} \frac{ L(\nu) }{h \nu} d\nu ,
\end{equation}

\noindent  where  $L(\nu)$  is  the AGN  monochromatic  luminosity  at
frequency $\nu$. The  \cite{Lusso2010} $L_\nu ({\rm 2 keV})-L_\nu (\rm 2500\,\AA)$
relation  is used  to convert  the  2-10\,keV X-ray  luminosity to  UV
monochromatic  luminosity at  2500\,\AA.  We  then extrapolate  to the
wavelength  range 227\AA\,  (4\,ryd)  and 910\AA\,  (1\,ryd) assuming  a
double power law for the AGN SED of the form 

\begin{equation}\label{eq_spectral_slopes} 
L(\nu) \propto \left\{
\begin{array}{ll}  \nu^{-0.5}   &  \mbox{(1100\,\AA  $<   \lambda  <$
2500\,\AA)} \\ \nu^{-1.7} & \mbox{$\lambda <$ 1100\,\AA} \\
\end{array} \right. \\
\end{equation}

\noindent  The   spectral  slopes  in  the  above   relation  are  from
\cite{vandenBerk2001}  and \cite{Telfer2002}.   At any  given redshift
the comoving density  of the hydrogen ionising rate  is then estimated
by integrating the X-ray luminosity function

\begin{equation} 
\mathcal{ N } = \int \phi(L_X,z) \, f_{esc}(L_X,z) \, \dot{n} \, {\rm dlog} L_X.
\end{equation}

\begin{figure}
\begin{center}
\includegraphics[height=0.8\columnwidth]{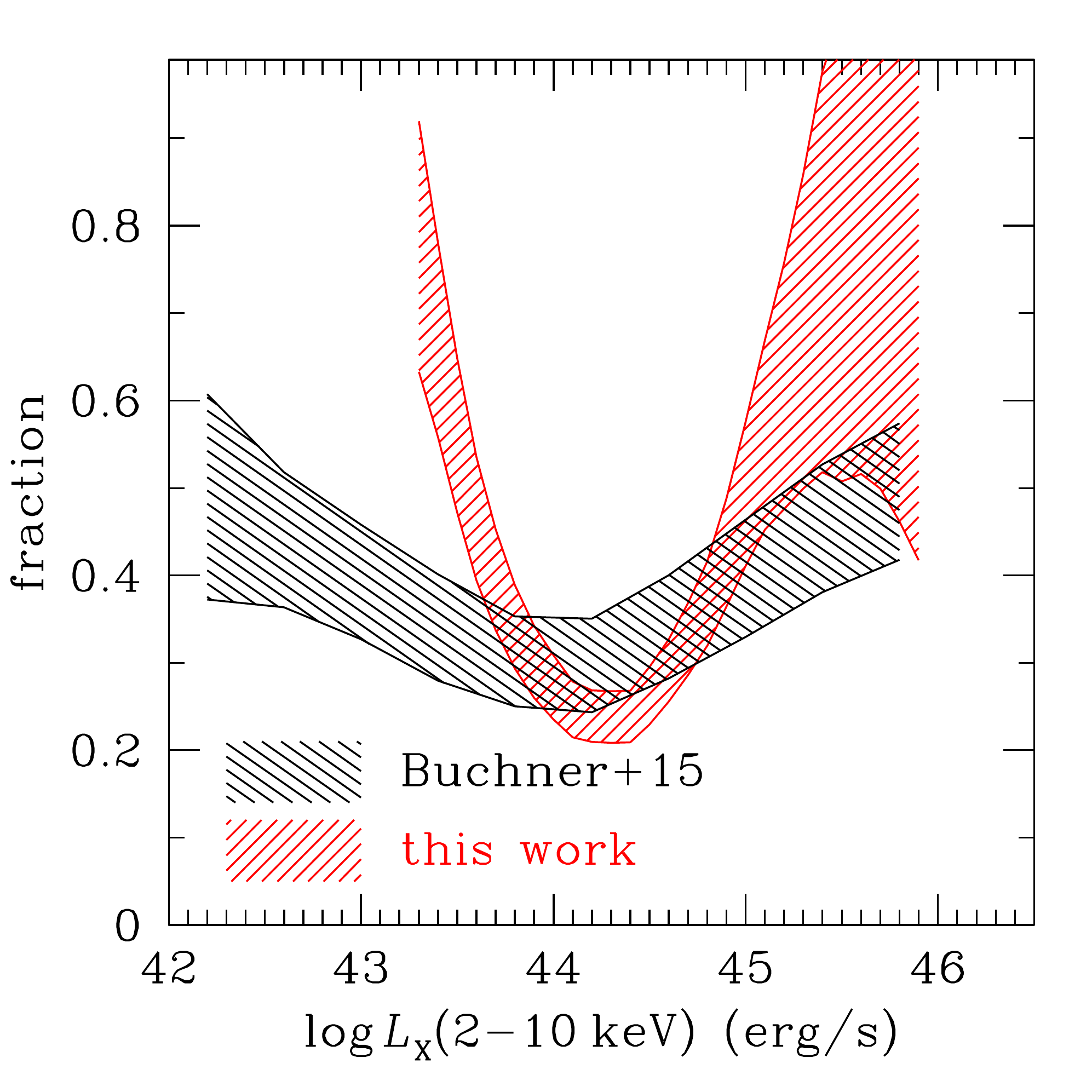}
\end{center}
\caption{Type-I AGN  fraction as a  function of X-ray  luminosity. The
red  shaded  region  is  the  ratio between  the  $z=3.2$  luminosity
functions of UV/optical QSOs from  Masters et al.  (2012) and our LDDE
parametrisation  for  X-ray AGN.  The  black  shaded  region is  $1  -
\mathcal{F}_{\rm obscured}$, where $\mathcal{F}_{\rm obscured}$ is the
obscured  AGN fraction  of Buchner  et  al.  (2015)  for the  redshift
interval $z=2.7-4$.  }\label{fig_typeI_fraction}
\end{figure}

\begin{figure}
\begin{center}
\includegraphics[height=0.8\columnwidth]{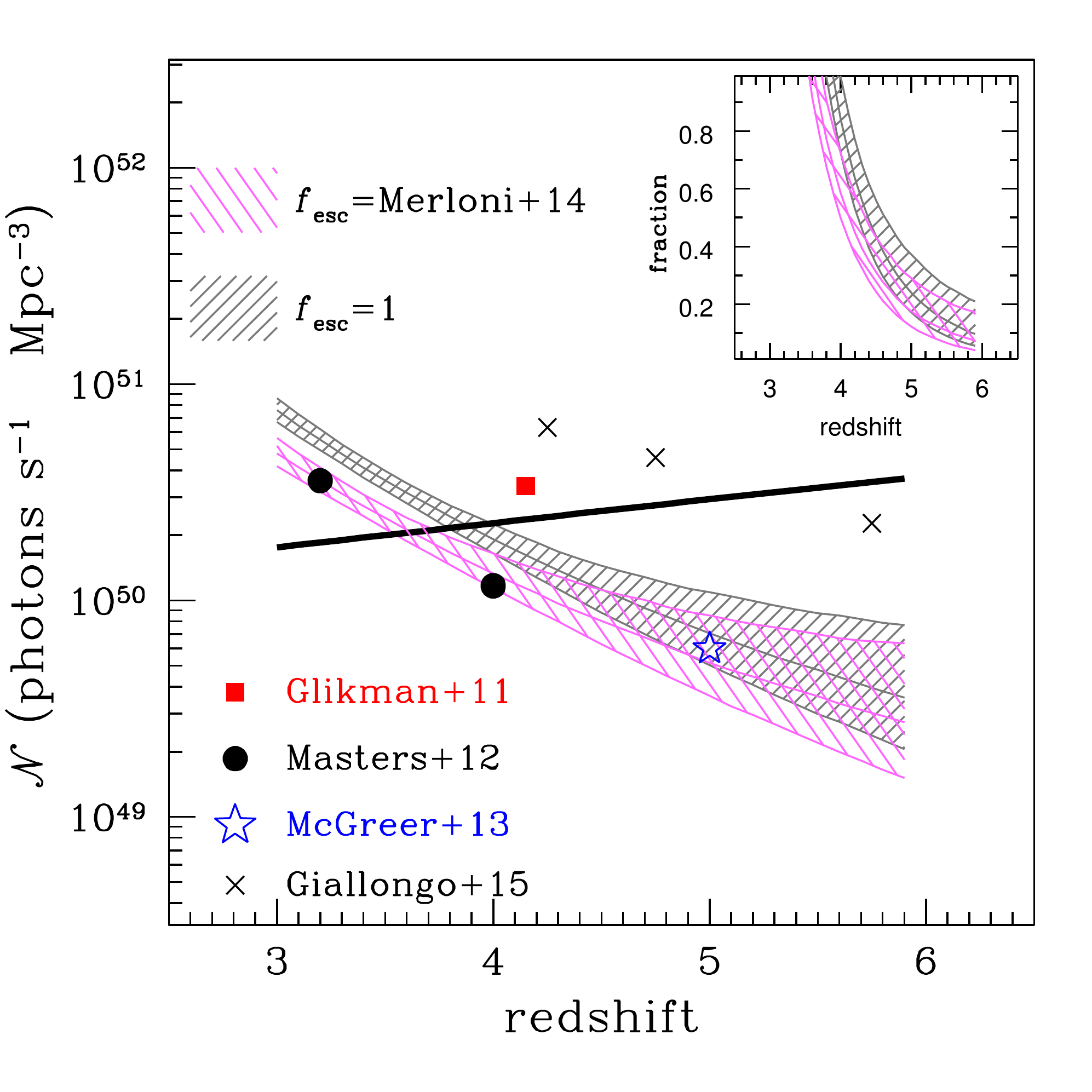}
\end{center}
\caption{Hydrogen  Ionising  photon  rate  density as  a  function  of
redshift.  The  shaded regions are  the constraints from  our analysis
using the  LDDE parametrisation for the X-ray  luminosity function and
under  different assumptions on  the escape  fraction of  AGN photons.
The  grey-shaded region assumes  an escaping  fraction of  unity, i.e.
ignoring  obscuration effects  close to  the supermassive  black hole.
The  pink-shaded region  assumes the  luminosity-dependent  Type-1 AGN
fraction of Merloni  et al.  (2014). We caution  that beyond $z=5$ the
shaded  curves  are  extrapolations.  The data  points  correspond  to
results in  literature.  The thick black line in the plot shows the photon rate density
required to keep the Universe ionised at any given redshift. The ratio
between the  shaded regions  and the black  line are presented  in the
inset  plot.}\label{fig_ionise}
\end{figure}

\noindent  The  LDDE  parametrisation  of  the  X-ray  AGN  luminosity
function  is adopted.   The  integration limits  are  set to  $L_X(\rm
2-10\,keV)=10^{42}$  and  $\rm 10^{46}\,  erg\,  s^{-1}$.  The  photon
escaping  factor, $f_{esc}(L_X,  z) $,  accounts for  the  fraction of
obscured  AGN, which  likely depends  on both  redshift  and accretion
luminosity. In these sources the ionising photons are absorbed locally
and  therefore do not  contribute to  the cosmic  ionisation radiation
field.  Our  analysis does  not constrain the  distribution of  AGN in
obscuration.    Additionally  at  redshift   $z>3$  there   are  still
discrepancies among different studies on the obscured AGN fraction and
its  dependence  on  luminosity  \cite[e.g.][]{Ueda2014,  Buchner2015,
Aird2015}.   For example,  in Figure~\ref{fig_typeI_fraction}  we find
evidence that  the type-I AGN fraction  and by proxy  the obscured AGN
fraction,  have  a  different  dependence on  luminosity  compared  to
relations  established at lower  redshift.  We  choose to  present our
baseline  results for  the  luminosity dependent  type-I AGN  fraction
determined by  \cite{Merloni2014}. In that study type-I  refers to AGN
with either blue UV/optical continua and/or broad emission lines. This
definition  is more  releavant to  the determination  of  the hydrogen
ionising photon  rates of AGN,  compared to e.g.  the  standrard X-ray
unobscured AGN definition, $\rm  N_H<10^{22} \, cm^{-2}$. Adopting the
\cite{Merloni2014}  relation for  $f_{esc}(L_X,z)$ also  allows direct
comparison  with previous  X-ray  studies that  also used  monotonically
increasing obscured  AGN fraction with decreasing  X-ray luminosity to
approximate the escape fraction of UV photons in high redshift AGN.
Figure~\ref{fig_ionise}  plots  $\mathcal{  N  }$  as  a  function  of
redshift for our  baseline model for the photon  escaping fraction. In
that figure we also place an upper limit in $\mathcal{ N }$ by setting
$f_{esc}=1$,  i.e.   the extreme  case  that  all  photons escape  and
contribute to  the hydrogen ionising radiation.  This  translates to a
net increase of  $\mathcal{ N }$ by a factor of  about 1.4 compared to
the \cite{Merloni2014} type-I AGN  relation for $f_{esc}$.  We further
explore how the results in Figure~\ref{fig_ionise} change if $f_{esc}$
is  approximated   by  X-ray  definitions  of   the  unobscured  ($\rm
N_H<10^{22}  \,  cm^{-2}$)  AGN  fraction.  We  adopt  the  luminosity
dependence   of  the   unobscured  X-ray   AGN  fraction   derived  by
\cite{Buchner2015}   in   the   redshift   interval   $z=2.7-4$   (see
Fig.~\ref{fig_typeI_fraction}).   This assumption  yields  photon rate
densities  that are  a factor  of about  1.7 smaller  compared  to the
baseline results  that use the \cite{Merloni2014}  type-I AGN fraction
as proxy of  $f_{esc}$.  For clarity these results  are not plotted in
Figure~\ref{fig_ionise}.

The constraints  above should be  compared to the minimum  photon rate
density required  to keep  the Universe ionised  at a  given redshift.
This is estimated from equation (26) of \cite{Madau1999} after scaling
it  to  our cosmology.   We  also  assume  that the  ionised  hydrogen
clumping  factor  in  that  relation,   which  is  a  measure  of  the
inhomogeneity of the medium, evolves with redshift as

\begin{equation} 
\mathcal{ C } = 1+ 43 \times z^{-1.71}. 
\end{equation}

\noindent   This  relation  is   based  on   cosmological  simulations
\citep{Pawlik2009}  and  is   adopted  by  \cite{Haardt_Madau2012}  to
synthesise  the evolving  spectrum of  the diffuse  UV/X-ray radiation
field.  Figure~\ref{fig_ionise}  plots the redshift  dependence of the
photon rate  density required to keep the  Universe ionised.  This
figure  shows that  for the  baseline model  \citep[i.e.][as  proxy of
$f_{esc}$]{Merloni2014}  the  AGN  contribution  to  the  photon  rate
density required  to keep the  Univese ionised decreases from  70\% at
$z=4$ to about 20\% at $z=5$. Assuming $f_{esc}=1$ the fractions above
translate  to upper  limits $<100\%$  at $z=4$  and $<30\%$  at $z=5$.
These numbers are in broad agreement with some previous studies on the
role   of   X-ray   AGN    in   the   ionisation   of   the   Universe
\citep{Barger2003_hiz, Haardt_Madau2012, Grissom2014}.  The decreasing
ionising photon  rate density fractions with increasing  redshift is a
direct consequence of the strong evolution of the X-ray AGN luminosity
function between redshifts $z=3$ and $z=5$.

 Finally in Figure~\ref{fig_ionise} the model constraints on the
ionising  photon rate  density at  $z=3-5$ from  our X-ray  sample are
compared  to previous  works based  on either  UV/optical QSO  surveys
\citep{Glikman2011, Masters2012, McGreer2013} or UV/X-ray selected samples
\citep{Giallongo2015}.   For the  calculation of  the ionising  photon
rate densities the  luminosity functions estimated in  these works are
integrated between absolute magnitudes $\rm M({\rm 1450\AA}) = -18$ and
$-28$\,mag    by    adopting    the   UV    spectrum    of    Equation
\ref{eq_spectral_slopes}.    Our    baseline   model    assuming   the
\cite{Merloni2014}   escaping   fraction   is  consistent   with   the
lower-normalisation UV/optical data points in Figure~\ref{fig_ionise}.

\begin{table*}
\caption{Parametric model best-fit parameters and uncertainties}\label{tab_xlf_fit}

\begin{tabular}{l cccc  cccc c}

\hline

Model & $\log K$  & $\log L_\star$ & $\gamma_1$ &  $\gamma_2$ &  $p$ & $q$  & $\beta$ & $\log_{10} \mathcal{Z}$ &  $\Delta \log_{10} \mathcal{Z}$ \\
      & ($\rm Mpc^{-3}$) & (erg/s)        &            &             &      &      &         &                   &                           \\

 (1) & (2) & (3) & (4) & (5) & (6) & (7) & (8) & (9) & (10) \\
\hline
PDE   & $-4.840_{-0.16}^{+0.13}$ & $44.33_{-0.12}^{+0.13}$ & $0.16_{-0.11}^{+0.15}$ & $2.03_{-0.20}^{+0.21}$ &  -- & $-6.59_{-0.93}^{+0.90}$ & -- & 9894.95 & 0.0\\

LDDE   & $-4.79_{-0.17}^{+0.14}$ & $44.31_{-0.11}^{+0.13}$ & $0.21_{-0.13}^{+0.15}$ & $2.15_{-0.21}^{+0.24}$ &  -- & $-7.46_{-1.12}^{+1.03}$ & $2.30_{-1.53}^{+1.60}$ & 9894.70 & --0.25\\

LADE   & $-4.78_{-0.17}^{+0.14}$ & $44.27_{-0.13}^{+0.14}$ & $0.18_{-0.12}^{+0.16}$ & $2.10_{-0.22}^{+0.26}$ &  $-0.85_{-0.17}^{+0.17}$ & $2.25_{-1.50}^{+1.47}$ & -- & 9893.62  & --1.33\\

PLE   & $-5.05_{-0.20}^{+0.14}$ & $44.42_{-0.14}^{+0.18}$ & $0.19_{-0.13}^{+0.17}$ & $1.73_{-0.15}^{+0.19}$ & $-4.98_{-0.80}^{+0.80}$ & -- & -- &  9890.93 &  --4.02\\

\hline

\end{tabular}
\begin{list}{}{}
\item 
Listed are  the best-fit  parameters for each  of the  four parametric
models considered in  this paper. The listed values are  the median of
the probability  distribution function  of each parameter.  The errors
correspond to  the 16th and  84th percentiles around the  median. The
columns  are: (1)  parametric model.  Models  are listed  in order  of
decreasing   Bayesian   evidence,   (2)   X-ray   luminosity   function
normalisation (see equation~\ref{eq_xlf}), (3) break luminosity of the
X-ray luminosity  function (see equation~\ref{eq_xlf}),  (4) faint-end
slope,  (5) bright-end  slope,  (6) density  evolution parameter  (see
equations~\ref{eq_pde},~\ref{eq_lade},~\ref{eq_ldde}), (7) luminosity
evolution parameters (see  equations~\ref{eq_ple},~\ref{eq_lade}), (8)
$\beta$ parameter  for the  LDDE model, (9)  base 10 logarithm  of the
Bayesian evidence for each model, (10) the difference between the
$\log_{10}
\mathcal{Z}$  of  each  model  and   the  PDE  that  has  the  highest
evidence. The  bayes factor of the  PDE model relative to  each of the
other three is $\exp(\Delta \ln \mathcal{Z})$.
\end{list}
\end{table*}

\begin{figure}
\begin{center}
\includegraphics[height=0.8\columnwidth]{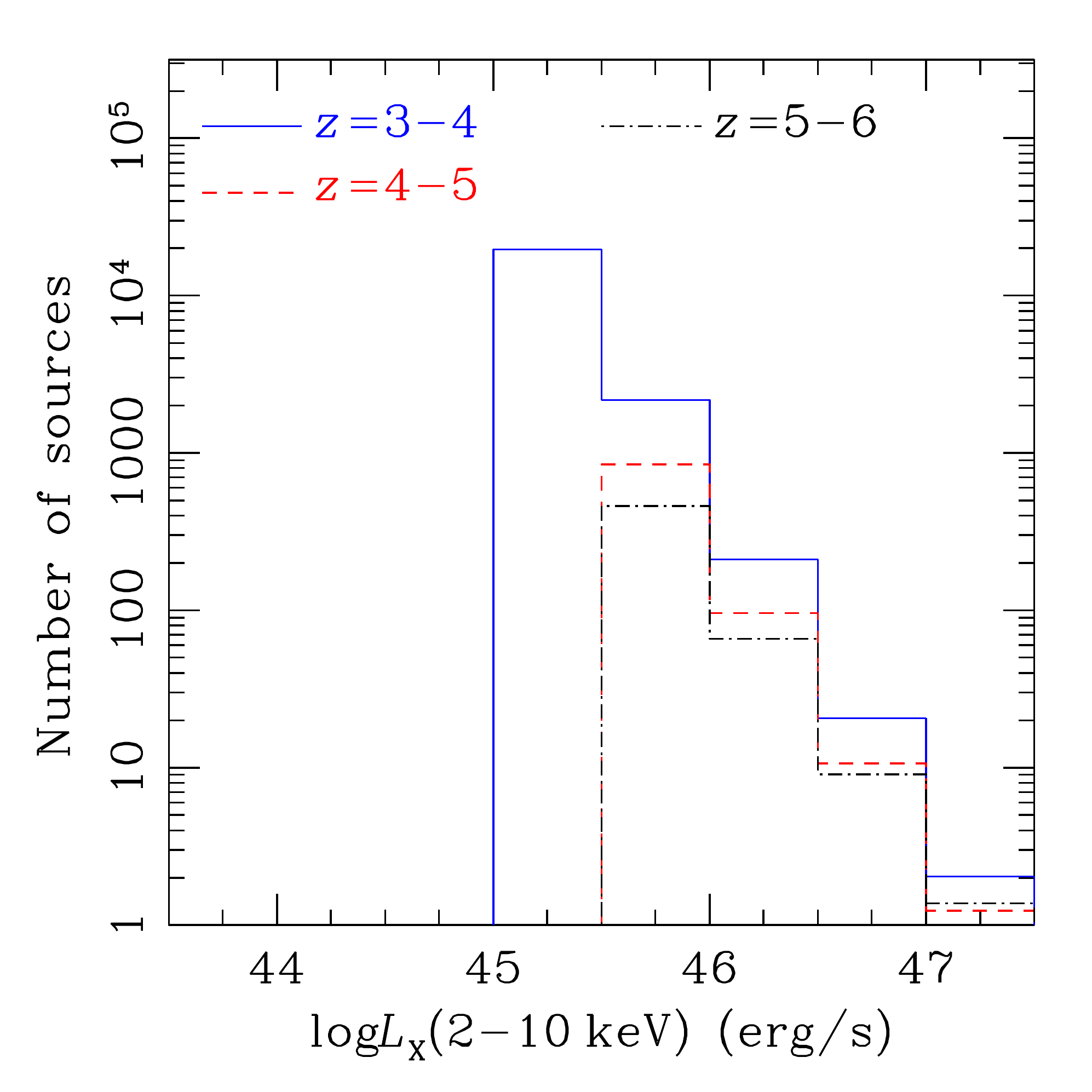}
\end{center}
\caption{Number of high redshift AGN  as a function of 2-10\,keV X-ray
  luminosity that  the eROSITA  4-year All Sky  Survey is  expected to
  detect. The LDDE parametrisation of the X-ray luminosity function is
  used  for the  predictions.  Results for  3  redshift intervals  are
  plotted,  $z=3-4$ (blue  solid),  $z=4-5$ (red  dashed) and  $z=5-6$
  (black dot-dashed). The predictions for  the latter redshift bin are
  extrapolations of the LDDE model.  The numbers are for 0.5\,dex wide
  luminosity  bins.}\label{fig_erosita}
\end{figure}

\section{Discussion}

In this  paper we  explore the evolution  of the AGN  X-ray luminosity
function in  the redshift interval  $z=3-5$ by combining  deep Chandra
surveys with  the wide-area and shallow XMM-XXL  sample.  This dataset
provides sufficient depth and volume to determine the space density of
AGN over 3\,dex  in luminosity, $\log L_X (\rm 2 -  10 \, keV) \approx
43 - 46$,  in the redshift intervals $z=3-4$  and $4-5$.  The analysis
methodology  we  develop takes  into  account  Poisson  errors in  the
determination   of  X-ray   fluxes   and  luminosities   as  well   as
uncertainties  in photometric  redshift  measurements. We  demonstrate
that the latter is critical for unbiased measurements of the AGN space
density  at  high  redshift.   Ignoring  photometric  redshift  errors
overestimates the X-ray luminosity function.  We also choose to follow
a non-parametric  approach and determine  the space density of  AGN in
luminosity and redshift bins.  This allows us to explore the evolution
of the AGN population  independent of model assumptions.  Additionally
when a  model parametrisation is applied  to AGN at  all redshifts the
fit  may be  driven  by  the redshift  and  luminosity intervals  that
contain  most  data.  This  can  introduce  biases  at high  redshift,
$z\ga3$,  where AGN samples  are typically  small as  a result  of the
rapid evolution  of the AGN  population as well as  survey sensitivity
and  volume  limitations.  We  account  for  this  potential issue  by
splitting   the   luminosity   function  into   independent   redshift
components, i.e. $z<3$, $z=3-5$.

We  confirm  previous  studies  for  a strong  evolution  of  the  AGN
population   in  the   redshift  interval   $z=3-5$  \citep{Brusa2009,
Vito2013,  Civano2012,  Kalfountzou2014}.   We  also  find  suggestive
evidence for  luminosity dependent  evolution of the  X-ray luminosity
function.  The space  density of AGN in the  luminosity interval $\log
L_X (\rm  2 -  10 \, keV)  \approx 43  - 45\, erg  \,s^{-1}$ decreases
faster  than  more  luminous  sources between  redshifts  $z=3.5$  and
$z=4.5$, albeit  at the 90\% significance level.   A similar evolution
pattern is  also observed in  the optical luminosity function  of QSOs
between  redshifts $z=3$ and  $z=4$ \citep{Masters2012}.   Our finding
can be  interpreted as evidence that  the formation epoch  of the most
powerful QSOs [$L_X  (\rm 2 - 10 \, keV)  \ga 10^{45}\, erg \,s^{-1}$]
precedes  that of  lower luminosity  systems. This  is similar  to AGN
downsizing  trends  established  at lower  redshifts  \citep{Ueda2003,
Hasinger2008,  Aird2010,  Miyaji2015}.   A  strong  evolution  of  the
faint-end of  the AGN luminosity function with  increasing redshift is
consistent with  the absence of X-ray  selected AGN at  $z\ga5$ in the
CANDELS  \citep{Grogin2011, Koekemoer2011}  subregion  of the  Chandra
Deep Field South \citep{Weigel2015}.   Extrapolating the LDDE model of
Table~\ref{tab_xlf_fit} we predict $<1$  AGN at redshift $z>5$ in that
field.

 Our analysis also places limits on the contribution of AGN to the
UV  photon  field  needed  to   keep  the  hydrogen  ionised  at  high
redshift. Using  empirical relations for  the type-1 AGN fraction  as a
function of  luminosity \citep{Merloni2014} we show  that AGN dominate
or at least  contribute a sizable fraction of  the required UV photons
to redshift $z\approx4$. At higher redshift the evolution of the X-ray
luminosity function  translates to a decreasing  contribution of X-ray
AGN to the UV photon field required to keep the hydrogen ionised.  The
extreme assumption of a photon  escaping fraction of unity for all AGN
sets  an upper limit  of 30\%  to the  contribution of  AGN to  the UV
photon rate  density required to  keep the hydrogen ionised  at $z=5$.
\cite{Barger2003_hiz}  use  multicolour  optical  data  in  the  2\,Ms
Chandra  Deep Field  North  \citep{Barger2003} and  conclude that  the
X-ray selected AGN  candidates at $z=5-6.5$ are too  few to ionise the
intergalactic  medium  at  those  redshifts.   \cite{Haardt_Madau2012}
estimated the  contribution of AGN  to hydrogen ionisation  rate using
the  \cite{Ueda2003}  X-ray luminosity  function  and AGN  obscuration
distribution.  They found that AGN do  not play an important role as a
source       of      ionising       photons       above      redshifts
$\approx4$.  \cite{Grissom2014} determine the  contribution of  AGN to
the ionisation of the hydrogen  in the Universe by taking into account
secondary  collisional  ionisations from  the  X-ray radiation.   They
extrapolate to high redshift ($z\ga6$) the \cite{Hiroi2012} hard X-ray
luminosity  function and  conclude that  AGN only  contribute  a small
fraction of the photon rate  densities required to ionise the Universe
at  these redshifts.   Our results  are  in agreement  with the  above
studies and  do not support claims for  a dominant role of  AGN to the
ionisation  of  the  hydrogen  in  the Universe  at  redshift  $z\ga4$
\citep{Glikman2011,   Giallongo2015}.   This  discrepancy   is  likely
related  to  the  way   different  groups  select  their  samples  and
subsequently  account for  this selection  in the  analysis.   It also
highlights  the need  for  further research  to  better constrain  the
impact    of    AGN   radiation    to    the    ionisation   of    the
Universe.  \cite{Glikman2011} estimate type-I  QSO space  densities at
$z\approx4$ that are  a factor of 3-4 higher  than those determined by
\cite{Masters2012}  or  \cite{Ikeda2011}   at  similar  redshifts  and
luminosities.     \cite{Giallongo2015}   combined   X-ray    and   HST
optical/near-IR data in the CANDELS GOODS-S region to identify optical
sources  with photometric  or spectroscopic  redshifts $z>4$  and then
study   their  X-ray   properties  following   methods   described  by
\cite{Fiore2012}.  Their  approach allows  them to identify  faint AGN
with X-ray luminosities as a low as $L_X \approx \rm 10^{43} \, erg \,
s^{-1}$  in the  redshift interval  $z=4-6.5$.  Strictly  speaking the
\cite{Giallongo2015} photon-rate  densities in Figure~\ref{fig_ionise}
are upper limits. The UV photon escape fraction is set to one, part of
the observed UV radiation of individual sources may be associated with
the  AGN host  galaxy, the  photometric redshift  uncertainties  of at
least some sources in the sample  are large. The increase of the X-ray
depth in the  CANDELS GOODS-S from 4 to 7\,Ms  (PI Brandt) region will
help better constrain the faint-end  of the AGN luminosity function at
high redshift and their role in the ionisation of the Universe.

Finally  the parametric  X-ray  luminosity functions  derived in  this
paper are  used to make  predictions on the  number of $z>3$  AGN that
eROSITA   \citep{Merloni2012}   surveys   will   detect.    Our   LDDE
parametrisation is  convolved with  the expected X-ray  sensitivity of
the 4-year eROSITA  All Sky Survey.  The number  of AGN in logarithmic
luminosity bins of size $\Delta \log L_X=0.5$ is plotted as a function
of 2-10\,keV luminosity  in Figure~\ref{fig_erosita}.  Predictions are
presented  for 3 redshift  intervals, $z=3-4$,  $4-5$ and  $5-6$. This
shows that  surveys by eROSITA  will provide tight constraints  on the
evolution  of  bright AGN  and  will allow  us  to  explore with  high
statistical significance the evidence for luminosity dependence of the
AGN population at  high redshift.  This however, would  also require a
dedicate  follow-up program to  identify high  redshift AGN  among the
eROSITA population.  High  multiplex spectroscopic facilities that are
able to  simultaneously observe  large number of  targets over  a wide
field of view  are essential for eROSITA X-ray  source follow-ups. The
SDSS/BOSS  spectrographs  \citep{Smee2013} at  the  Apache Point  SDSS
telescope \citep{Gunn2006}  and in  the future the  ESO/4MOST facility
\citep[4-metre                Multi-Object               Spectroscopic
Telescope,][]{deJong2014_4most}   are   well   suited  for   follow-up
observations of the eROSITA sky.

\section{Conclusions}

X-ray  data from  Chandra deep  surveys and  the  shallow/wide XMM-XXL
sample are combined  to explore the evolution of  the X-ray luminosity
function at high redshift, $z=3-5$.  Our analysis accounts for Poisson
errors  in the  calculation  of  fluxes and  luminosities  as well  as
photometric  redshift uncertainties.   We  also show  that the  latter
point  is crucial for  unbiased AGN  space density  measurements.  The
sample used  in the paper consists  of nearly 340  sources with either
photometric  (212) or  spectroscopic  (128) redshift  in the  redshift
range $z=3-5$.   The luminosity  baseline of the  sample is  $L_X (\rm
2-10\,keV) \approx 10^{43}-10^  {46} \,erg \,s^{−1}$ at $z  > 3$.  Our
main findings are

\begin{itemize}

  \item  the  AGN population  evolves  strongly  between the  redshift
intervals $z=3-4$ and $z=4-5$

  \item there  is evidence,  significant at the  90\% level,  that the
amplitude of  the AGN evolution  depends on X-ray  luminosity. Sources
with  luminosities $L_X(\rm  2 -  10\,keV) <10^{45}\,  erg  \, s^{-1}$
appear to evolve faster than brighter ones.

  \item the faint-end slope  of UV/optical QSO luminosity functions is
steeper than that  of the X-ray selected AGN  samples. This implies an
increasing fraction of type-I  AGN with decreasing X-ray luminosity at
$z>3$.

  \item X-ray AGN may dominate or at least contribute substantially to
the UV photon  rate density required to keep  the Universe ionised
to $z=4$.  At higher redshift the  contribution of AGN  to UV hydrogen
ionising field decreases.

\end{itemize}

\section{Acknowledgements}

This work has  made use of the Rainbow  Cosmological Surveys Database,
which  is operated  by the  Universidad Complutense  de  Madrid (UCM),
partnered  with the  University of  California Observatories  at Santa
Cruz  (UCO/Lick,UCSC).   This work  benefited  from  the {\sc  thales}
project 383549  that is jointly funded  by the European  Union and the
Greek  Government in the  framework of  the programme  ``Education and
lifelong  learning''. Funding for  SDSS-III has  been provided  by the
Alfred  P.  Sloan  Foundation,  the  Participating  Institutions,  the
National Science Foundation, and  the U.S. Department of Energy Office
of Science.  The SDSS-III web site  is http://www.sdss3.org/. SDSS-III
is  managed   by  the   Astrophysical  Research  Consortium   for  the
Participating Institutions of the SDSS-III Collaboration including the
University of  Arizona, the Brazilian  Participation Group, Brookhaven
National  Laboratory,   Carnegie  Mellon  University,   University  of
Florida,  the  French Participation  Group,  the German  Participation
Group, Harvard  University, the Instituto de  Astrofisica de Canarias,
the Michigan State/Notre  Dame/JINA Participation Group, Johns Hopkins
University,   Lawrence  Berkeley   National  Laboratory,   Max  Planck
Institute for Astrophysics,  Max Planck Institute for Extraterrestrial
Physics, New Mexico State  University, New York University, Ohio State
University, Pennsylvania  State University, University  of Portsmouth,
Princeton University,  the Spanish Participation  Group, University of
Tokyo,  University  of  Utah,  Vanderbilt  University,  University  of
Virginia, University of Washington, and Yale University.

\bibliography{/home/age/soft9/BIBTEX/mybib}{}
%\bibliography{/soft9/BIBTEX/mybib}{}
\bibliographystyle{mn2e}

\end{document}